\documentclass[fleqn,usenatbib]{mnras}

\usepackage{newtxtext,newtxmath}

\usepackage[T1]{fontenc}

\DeclareRobustCommand{\VAN}[3]{#2}
\let\VANthebibliography\thebibliography
\def\thebibliography{\DeclareRobustCommand{\VAN}[3]{##3}\VANthebibliography}

\usepackage{tikz,xcolor,hyperref}

\usepackage{graphicx}	
\usepackage{amsmath}	
\usepackage{gensymb}
\usepackage{longtable}
\usepackage{anyfontsize}

\newcommand{\qvar}{Q_\text{var}}
\newcommand{\tsf}{\text{T}_\text{SF}}
\newcommand{\hb}{H$\beta$}
\newcommand{\oiii}{[O{\footnotesize III}]}

\definecolor{lime}{HTML}{A6CE39}
\DeclareRobustCommand{\orcidicon}{%
    \begin{tikzpicture}
    \draw[lime, fill=lime] (0,0) 
    circle [radius=0.16] 
    node[white] {{\fontfamily{qag}\selectfont \tiny ID}};
    \draw[white, fill=white] (-0.0625,0.095) 
    circle [radius=0.007];
    \end{tikzpicture}
    \hspace{-2mm}
}

\newcommand{\orcidChrisO}{\href{https://orcid.org/0000-0003-0017-349X}{\orcidicon}}
\newcommand{\orcidChrisW}{\href{https://orcid.org/0000-0002-4569-016X}{\orcidicon}}
\newcommand{\orcidJT}{\href{https://orcid.org/0000-0003-2858-9657}{\orcidicon}}
\newcommand{\orcidNA}{\href{https://orcid.org/0009-0005-7553-049X}{\orcidicon}}
\newcommand{\orcidRW}{\href{https://orcid.org/0000-0002-5325-2709}{\orcidicon}}
\newcommand{\orcidJH}{\href{https://orcid.org/0000-0001-8359-2328}{\orcidicon}}
\newcommand{\orcidSL}{\href{https://orcid.org/0000-0001-9372-4611}{\orcidicon}}

\title[Discovering CLAGN from 6dFGS and ATLAS]{Discovering Changing-Look AGN in the 6dF Galaxy Survey using ATLAS Light curves} 

\author[Amrutha et al.]{
Neelesh Amrutha,$^{1}$\thanks{E-mail: neelesh.amrutha@anu.edu.au}\orcidNA
Christian Wolf,$^{1,2}$\orcidChrisW
Christopher A. Onken,$^{1,2}$\orcidChrisO
Wei Jeat Hon,$^{3}$\orcidJH
\newauthor{}
Samuel Lai,$^{1}$\orcidSL
John L. Tonry$^{4}$\orcidJT
and Rachel Webster$^{3}$\orcidRW
\\
$^{1}$Research School of Astronomy and Astrophysics (RSAA), Australian National University, Canberra ACT 2611, Australia\\
$^{2}$Centre for Gravitational Astrophysics (CGA), Australian National University, Building 38 Science Road, Acton ACT 2601, Australia \\
$^{3}$School of Physics, University of Melbourne, Parkville, Victoria 3010, Australia \\
$^{4}$Institute for Astronomy, University of Hawaii, 2680 Woodlawn Drive, Honolulu, HI 96822-1897, U.S.A. \\
}

\date{Accepted XXX. Received YYY; in original form ZZZ}

\pubyear{2024}

\begin{document}
\definecolor{asparagus}{rgb}{0.53, 0.66, 0.42}
\label{firstpage}
\pagerange{\pageref{firstpage}--\pageref{lastpage}}
\maketitle

\begin{abstract}
Changing-Look Active Galactic Nuclei (CLAGN) are characterised by extreme variations in line emission over short timescales, mostly affecting broad \hb\ lines. While a few hundred CLAGN are known, a complete sample of turn-on CLAGN is still elusive. Here, we present a search for turn-on CLAGN in a complete sample of galaxies, using archival spectra and recent light curves. We obtained light curves from the Asteroid Terrestrial Impact Last Alert System (ATLAS) for 16\,232 emission line galaxies, including both star-forming and active galaxies, at $z<0.1$ with spectra from the Six-degree Field Galaxy Survey (6dFGS). We first establish typical variability behaviour for different AGN types, as recorded between 2001 and 2009, and then select outliers from the bulk behaviour as CLAGN candidates. We obtain new spectra for the candidates and identify 12 new turn-on CLAGN (appearing broad \hb\ line) and 19 new turn-off CLAGN (disappearing broad \hb\ line). We may have missed AGN that changed and reverted their state over the 15$-$20 years since 6dFGS spectra were taken, and thus our CLAGN rates of 1.7\% for turn-on and 9.6\% for turn-off are lower limits. The turn-on rate is naturally much lower as the type 1.9/2 sample is dominated by obscured AGN due to orientation, which are not expected to change. However, the number of turn-on (27) and turn-off (24) CLAGN we find are similar, suggesting that our parent AGN sample is reasonably complete in our search volume at $z<0.1$.
\end{abstract}

\begin{keywords}
methods: observational – galaxies: active - galaxies: Seyfert – accretion, accretion discs
\end{keywords}



\section{Introduction}

The unification model of Active Galactic Nuclei (AGN) \citep{antonucci_1993, urry_padovani_1995} provides a simple explanation for the "zoo" of observed AGN properties. This model hinges on the orientation of the AGN and its impact on the visibility of the Broad Line Region (BLR). In recent years, the field of time-domain astronomy has become increasingly data-rich, leading to the discovery of a growing population of AGN that exhibit strong changes in the strength of their Broad Emission Lines (BELs) over relatively short timescales of a few months or years \citep{tohline_osterbrock_1976,zeltyn_trakhtenbrot_2022}. These AGN are called Changing-Look (CL) AGN \citep{matt_guainazzi_2003}. Of course, a static unification model cannot explain phenomena that allow for these drastic changes in the observed BLR, as the orientation of an AGN cannot rapidly change within the observed CL timescales.

The phenomena associated with CLAGN can be categorised into the two primary classes of Changing-Obscuration AGN (COAGN) and Changing-accretion State AGN (CSAGN) \citep{ricci_trakhtenbrot_2022}. While both of these CL classes manifest as a dis/-appearance of UV-optical continuum variability corresponding to the dis/-appearance of broad lines \citep{lopez-navas_arevalo_2023}, the driving mechanism of these apparent changes are different. COAGN are typically identified through multi-epoch X-ray observations, where a line-of-sight column density of $N_H = 10^{22} \text{ cm}^{-2}$ has been suggested as a critical threshold for obscuration \citep{koss_trakhtenbrot_2017,ricci_trakhtenbrot_2017}. This column density threshold is also applicable in the optical wavelengths \citep{merloni_bongiorno_2014}, with higher column density obscuring the broad lines. CSAGN are extremely variable objects, where rapid changes in the accretion rate vary the strength of the broad lines. Assuming a standard thin disc model \citep{shakura_sunyaev_1973}, such extreme variability changes are expected to occur on the accretion disc viscous timescale, on the order of $>$10$^3$ years for an isolated disc, while observed CS timescales are much shorter. CSAGN currently dominate the literature, as the obscuration phenomenon has been ruled out for most CLAGN cases through a variety of means, including observations in polarised light \citep{hutsmekers_gonzalez_2017}, mid-IR variability amplitude and optical time lag expected for dust-reprocessing \citep{sheng_wang_2017}, examination of unrealistic timescales of passing dust clouds \citep{lamassa_cales_2015,zeltyn_trakhtenbrot_2022}, or directly through the presence of unobscured AGN features in the X-ray spectrum \citep{noda_done_2018}. In particular, \citet{zeltyn_trakhtenbrot_2022} observe a case where transient obscuration would best explain the features of CL event, but the timescale of a few months is a factor of $\sim$100 shorter than expected for a cloud to pass across the line of sight. Hereafter, CLAGN includes both COAGN and CSAGN unless specified otherwise.

While the initial discoveries of CLAGN were largely serendipitous \citep{tohline_osterbrock_1976,shappee_prieto_2014,denney_de_rosa_2014, lamassa_cales_2015}, nearly 1000 CLAGN have been identified to date, primarily as a result of systematic searches conducted over the past decade \citep[eg.,][]{macleod_ross_2016,macleod_green_2019,yang_wu_2018_clagn,graham_ross_2020_csq,temple_ricci_2023_clagn,guo_zou_2024_clagn2,guo_zou_2024_clagn1,zeltyn_trakhtenbrot_2024}. These discoveries were typically achieved through the comparison of multi-epoch spectral data obtained from extensive surveys such as the Sloan Digital Sky Survey (SDSS) or the BAT AGN Spectroscopic Survey (BASS). In cases where multi-epoch spectra are not available, targeted selection methods use expected photometric behaviour of BEL AGN in terms of colour \citep{wolf_golding_2020, hon_wolf_2022_skymapper_colors} or light curve variability \citep{lopez_martinez_2022} to identify a small sample of CLAGN candidates in order to observe a second epoch spectral AGN type.

Optical spectra of AGN are traditionally categorised into two primary types: type 1 AGN show both broad and narrow emission line components, while type 2 AGN exhibit only narrow line emission components. The lack of broad lines in type 2 AGN is explained by either an obscured BLR or a "true" lack of a BLR due to a low accretion rate. The BLR can be obscured due to the orientation of the host galaxy, by dust within the nuclear region of the host, or galaxy scale material \citep{hickox_alexander_2018}. The broad lines from these hidden type 1 AGN can be observed in polarised light \citep{tran_2001,tran_2003}. "True" type 2 AGN do not contain a BLR, hidden or otherwise, with an unobscured line-of-sight. For example, type 2 AGN which show CS behaviour are "true" type 2 where the BLR is not hidden by orientation. The primary type 1 and 2 classification suggests a binary division, although a broad range of intensity ratios is observed between the broad and narrow emission components. To account for this diversity, an intermediate sub-type classification system has been established by \citet{osterbrock_1981}, denoted as types 1.2, 1.5, 1.8, and 1.9. In this sub-type sequence, the strength of the broad Balmer lines decreases from type 1 to type 2, until higher-order Balmer lines become weak or undetectable in type 1.8 and 1.9 objects.

Accretion discs in AGN are strong sources of UV-optical emission, which are known to exhibit brightness variations on timescales of weeks to years \citep{giveon_maoz_1999,vandenBerk_wilhite_2004,wilhite_brunner_2008,ruan_anderson_2014,dexter_begelman_2019}. When AGN change their state between non-accreting and accreting, we then also expect their optical continuum emission to change between non-variable and variable light curves \citep{lopez-navas_sanchez-saez_2023}; the same change is expected when the variable accretion disc gets temporarily obscured by dust. Hence, AGN types can be estimated from light curves, although they are only reliably determined from (and are defined by) emission-line diagnostics in their spectra.

While the sample of known CLAGN is ever increasing, targeted searches of CLAGN particularly in the process of changing would be required to further constrain timescales and mechanisms of CL events. Most CLAGN identification studies search within a known BEL AGN sample, particularly quasar samples, and thus would exclude type 2 AGN with a hidden BLR. Searching within a type 2 sample would include both "true" type 2 AGN which can exhibit CL behaviour and hidden type 1 AGN where CL events would not be identified due to obscuration. As such, CLAGN searches in a type 2 AGN sample are rare. Furthermore, only a few large sample studies of type 2 AGN are available \citep[eg.,][]{kauffmann_heckman_2003,reyes_zakamska_2008,vietri_garilli_2022}, and even fewer targeted multi-epoch CLAGN searches on a complete type 2 AGN sample \citep[eg.,][]{yang_green_2024_arxiv}.

This work aims to find CLAGN from a complete galaxy sample at $z<0.1$ using the optical variability of light curves. The principal idea behind our search for CLAGN is to combine historic AGN types from a large spectroscopic survey with current-epoch light curves from a large photometric monitoring program. In this paper, we use historic spectra from the 6dF Galaxy Survey \citep[6dFGS,][]{jones_saunders_2004_6dfgs,jones_read_2009_6dfgs}, which contains spectra of over 100,000 low-redshift galaxies across most of the Southern sky observed between the years 2001 and 2009. We combine this rich source of nearby galaxies with light curves from the NASA Asteroid Terrestrial Impact Last Alert System (ATLAS) \citep{tonry_denneau_2018_atlas}; ATLAS has been monitoring more than three-quarters of the entire sky since 2016.

This paper is organised as follows: in Section~\ref{sec:Data}, we present the sample and data used in this project, and the steps involved in cleaning the data set. Section~\ref{sec:Methods} details the methods used for classifying light curves into variable and non-variable, lays out the results of this classification and identifies CLAGN candidates for confirmation with follow-up spectroscopy. Section~\ref{sec:spec-follow-up} then presents the results of the follow-up and Section~\ref{sec:discussion} discusses these findings. Throughout the paper, we use AB magnitudes and adopt a flat Lambda cold dark matter cosmology with $\Omega_\Lambda = 0.7$ and $H_0 = 70$~km~s$^{-1}$~Mpc$^{-1}$.

\section{Data}\label{sec:Data}

The 6dF Galaxy Survey (6dFGS) is a redshift survey with a primary sample selected using near-infrared (NIR) $K$-band magnitudes from the Two Micron All Sky Survey Extended Source Catalogue \citep[2MASS;][]{jarrett_chester_2000}. 6dFGS contains 125,071 galaxies complete to $K<12.65$ mag with a median redshift of $z\sim0.053$ covering the Southern Sky with $|b|\geq10\degree$. Using a complete sample of galaxies, we expect to find a complete sample of AGN, by identifying AGN spectra from 6dFGS. This includes an unprecedented sample of type 2 AGN that is complete at $z<0.1$ at the K$<12.65$ mag limit. 

We fit all 6dFGS spectra using redshifts provided by the 6dFGS catalogue to identify emission line peaks with a significance threshold greater than three standard deviations. Emission line galaxies are separated from other galaxies by the strong \oiii\ and H$\alpha$ emission. These emission line galaxies are fit between $4000-7000$\AA\ with narrow Gaussian components for H$\alpha$, \hb, \oiii\ and [N{\footnotesize II}] and a broad Gaussian component for the Balmer lines. The upper limit on the FWHM for the blended narrow lines (H$\alpha$, \hb, and [N{\footnotesize II}]) is set to 700 km s$^{-1}$, while the unblended \oiii\ line is allowed to go up to 1200 km s$^{-1}$, which is also the lower limit for the FWHM of the broad component. AGN and star-forming (SF) galaxies in 6dFGS are identified by their emission lines and are differentiated by their narrow-line ratios informed by BPT diagrams \citep[eg.,][]{kewley_dopita_2001,stasinska_cid_2006}. Around 3\,600 composite galaxies that are ambiguous on the BPT diagram with no broad features are omitted from this study after a visual examination of the spectra. 

We thus generate an AGN catalogue from 6dFGS spectra, hereafter called 6dFAGN. We also generate a more carefully selected list of BEL AGN with the sub-types identified from the \hb/\oiii\ ratio (Hon et al., submitted). The role of SF galaxies is to inform the noise level in the light curves, as these galaxies have no detectable AGN and will thus be virtually all non-variable, except for rarely occurring supernovae and Tidal Disruption Events \citep[TDEs;][]{gezari_tde_2021}. For this work, we limit the sample to $z<0.1$, where 6dFAGN is a complete sample of AGN in massive galaxies down to a stellar mass of $\log\mathrm{M/M}_\odot \sim 10.4$ using the K-band mass-to-light ratio from \cite{mcgaugh_schombert_2014_m_lks_ratio} at the median redshift of the 6dFGS sample, $z\sim0.053$. The $z<0.1$ threshold corresponds to a redshift where the $\lambda 5007$\AA\ \oiii\ emission line approaches the $\lambda 5577$\AA\ night-sky emission line; therefore, we avoid the interference of the strong sky line with any characterisation of the AGN involving the \oiii\ emission line.

In some cases, scattered light within the spectrograph has contaminated regions occupied by neighbouring spectra on the detector. Consequently, this cross-talk between fibres creates fake type 1 AGN with broad lines from light leaked from true type 1 AGN. These are identified by multiple sets of emission lines at different redshifts along with consecutive spectral IDs within 6dFGS and are subsequently removed from our CLAGN search sample. In addition to cross-talk spectra, we clean the 6dFGS sample based on the quality of spectra where the continuum and emission lines cannot be reliably fit, remove any neighbours within 4" that could contaminate the light curve using {\it Gaia} DR3 \citep{gaia_dr3}, and also remove high Galaxy reddening \citep{irsa_dust_schlegel_1998} with an E(B$-$V)$>$0.3 cut-off, thereby removing 119 AGN spectra which is a small fraction of our final sample. 

While the sky area of 6dFGS is limited to South of the celestial equator, a large part of it lies in the footprint of the Northern-focused Asteroid Terrestrial Impact Last Alert System (ATLAS) operated by NASA. The system was first operational in 2015 with a single 0.5-meter telescope. A second telescope was added in 2017, and two more were added in early 2022. The first two are located at Haleakala and Mauna Loa observatories in Hawaii, allowing observations of the Northern sky down to a declination $-45\degree$. The newer two telescopes are located at Sutherland Observatory in South Africa and El Sauce Observatory in Chile, allowing observations of the Southern sky. However, 6dFGS sources with ATLAS data only from the Southern telescopes are omitted from our search sample, due to the limited period of available light curve data. Observations are conducted with 30-second exposures and a 28.9 square degrees field of view. The telescopes have a pixel scale of 1.86” with mosaic dimensions of 10,560 $\times$ 10,560 pixels. The depth of ATLAS is sufficient to monitor all galaxies in this nearby sample that are part of the 6dFGS main sample, whether they are AGN or not. Before the addition of the two new telescopes, the telescopes scanned the sky four times every two days. ATLAS provides difference-image photometric light curves in two broad passbands: cyan ($c$, 420-650 nm) and orange ($o$, 560-820 nm). For our analysis, we use the difference-image light curves in flux density units. The combined sample in this study covers $-45\degree < \delta < 0\degree$ and $|b|\geq10\degree$, with light curve data from March 2017 and 6dFGS spectra for 2,698 AGN and 13,534 star-forming galaxies at $z<0.1$.

\section{Methods}\label{sec:Methods}

In this section, we describe how we characterise variability in the light curves, and how we use the statistical noise levels in non-variable galaxies to tune a selection method for truly variable objects. This tuning involves an evaluation of completeness and contamination as a function of the variability threshold.

\subsection{Light curve cleaning}
We clean the light curves for any erroneous measurements that could affect our variability analysis.
We performed an iterative sigma-clipping of the errors in the light curve measurements with a $3\sigma$ threshold and removed measurements with large errors across the entire light curve period. We found a generic first-order error cut-off at 45 $\mu$Jy clipped about 15\% of the measurements. Increasing the error threshold increases the light curve noise, without enhancing AGN variability. We then stacked the $o$-band measurements into 7-day bins, with nine measurements per stack on average. These intervals were chosen to strike a balance between noise suppression and the preservation of genuine variations. Due to the limited frequency of $c$-band measurements, which were exclusively observed during the dark time of the lunar cycle, we stacked the $c$-band data into bins centred around new-moon dates with eleven measurements per stack on average. Since the $o$-band measurements are much better sampled and the number of $c$-band measurements varies greatly between light curves, we ignore $c$-band measurements in our quantitative analysis. However, we include them in our plots for visual inspection, noting that the bluer $c$-band is more sensitive to AGN variability relative to the $o$-band.

Each stack is represented by the median value of measurements in the stack, as the mean is sensitive to outliers in a small sample. Given our goal to characterise light curve variability, we can retain some information about the variability within the stacked period by setting the inter-quartile range (IQR) as the confidence interval of the stack. As a result, stacks with fewer than three measurements were excluded. For stacks with less than five measurements, quantiles were computed by dividing the range into linear intervals and interpolating percentile values.

In rare cases, coincident measurements may lead to an IQR that is smaller than the expected measurement error of the stack and would then suggest an unrealistically small confidence interval. To prevent this, we compute the error of the stack points from the following:
\begin{equation}
    X_\mathrm{err}= \text{max}\left(\frac{\text{IQR}}{2},\text{ mean measurement error}\right). \label{eq:err}
\end{equation}
Our decision to exclude the $1/\sqrt{N}$ term from the mean measurement error calculation was to ensure that the stacked errors were in line with the variability of measurements within each stacking period, mirroring the behaviour of the IQR. The stacked measurements were further refined by following a similar recipe of sigma-clipping errors and removing measurements with large errors. The final stacked measurements after clipping result in a maximum error of 50 $\mu$Jy for the $o$-band and 35 $\mu$Jy for the $c$-band. A final condition in our light curve cleaning was the exclusion of any light curves with fewer than 90 week-long stacks over six years.

ATLAS has reported that the reference image used for difference photometry, referred to as the \textit{wallpaper}, was updated on two occasions, at MJD 58417 (October 26, 2018) and MJD 58882 (February 3, 2020). These updates may introduce step-like discontinuities in the light curve data at the specified dates when the light curve exhibits significant variability before the wallpaper update. For this study, we have chosen not to correct such discontinuities, because they only increase the level of measured variability in variable light curves, while non-variable light curves remain unaffected. However, for the plots in the paper, wallpaper jumps are corrected by downloading the raw light curves and matching the difference in average of the entire light curve before and after the wallpaper update.

\subsection{AGN classification}
The 6dFGS observations use a large 6.7" fibre. At $z=0.05$, this corresponds to an angular size of 6.6 kpc, or 12.4 kpc at $z=0.1$. Thus, the fibre includes a significant stellar contribution, including narrow emission lines from star formation. For this work, we adopt the classification system based on the \hb/($\lambda$5007\AA\oiii) line ratio, outlined in \citet{winkler_1992}. We consider solely the broad component of \hb, expected to emanate from the vicinity of the accretion disc, as opposed to the total \hb\ luminosity, because the narrow component may also originate from star formation within the host galaxy. The number of AGN in each sub-type bin in our search sample after cleaning is listed in Table \ref{tab:agn_count}. 

Throughout the paper, we choose to combine type 1.9 and 2 objects into a single AGN type, which differ only by the presence or absence of a broad component in the H$\alpha$ emission line. While a strong broad H$\alpha$ line would warrant a type 1.9 label, we believe that a robust distinction between type 1.9 and type 2 is inherently difficult, especially in light of host dilution in large-aperture spectra. The success of any such distinction will depend on data quality and host dilution, which depend on distance and AGN luminosity. 

For the BEL AGN sample, the median $\lambda5007$\AA\ \oiii\ emission luminosity is L$_\mathrm{[OIII]} \approx 10^{41.2}$ erg s$^{-1}$ which corresponds to a $2-10$ keV X-ray luminosity of L$_{2-10} \approx 10^{42.8}$ erg s$^{-1}$ using the relation found in Equation 4 of \citet{hasinger_2008} \citep[derived from][]{netzer_mainieri_2006}. This corresponds to a type 2 AGN fraction of 75\%, using the type 2 fraction and X-ray luminosity relation in \citet{hasinger_2008}. However, our type 1.9/2 fraction is 86\%, as narrow emission lines are much easier to detect compared to weak but broad lines, therefore including lower-luminosity type 2 AGN in our sample.

Using the 5100\AA\ monochromatic AGN luminosity from our BEL AGN sample, along with a bolometric correction from \citet{shen_richards_2011}, we estimate a median luminosity $\mathrm{L}_\mathrm{bol} \approx 10^{44.4}$ erg s$^{-1}$ for our sample. Note that the 6dFGS spectra are not flux calibrated and the reported luminosity values in this section are estimated from fits to recent follow-up spectra of our BEL AGN sample as explained in Section \ref{sec:spec-follow-up}.

\subsection{Variability measure}
A quantitative measure of an object's variability can be obtained by subtracting the measurement noise from the light curve's variance, resulting in the excess variance of the light curve. However, the variance of the intra-stack measurements of BEL AGN in our sample is comparable to the measurement noise, allowing for a higher tolerance for the excess variability compared to SF galaxies. This reduces the variability contrast between SF and BEL AGN. In the raw, unstacked light curves, excess variability is dominated by photometric outliers. For BEL AGN, distinguishing whether these outliers are due to actual variability or measurement noise is difficult.

To separate variable from non-variable objects, we can estimate a simple variability quantity such as the root-mean-squared error (RMSE) of the light curve. However, the RMSE is sensitive to outliers in the light curve. Furthermore, the RMSE does not account for the brightness dependence of instrumental noise and variability amplitude. Additionally, we restrict the light curve data to just the $o$-band as it is much better sampled than the $c$-band measurements. Considering these factors, we define a variability metric that can be readily corrected for magnitude-dependent measurement noise:
\begin{equation}
    \qvar = \log\left(\frac{R_{98-2}^o}{\langle X_\mathrm{err} \rangle}\right),
\end{equation}
where $R_{98-2}^o$ is the 2nd to 98th percentile range of the $o$-band light curve. $\langle X_\mathrm{err} \rangle$ is the mean of the errors given by Equation \ref{eq:err} which down-weights in $\qvar$ large noise amplitudes with relatively small mean IQR. To scale the measured variability by the expected brightness-dependent noise, we use $r$-band magnitudes ($r_\text{psf}$) from the SkyMapper Southern Sky survey \citep[SMSS DR3;][]{wolf_onken_2018_smss, onken_wolf_2019_smss_dr2} which significantly overlaps with the $o$-band. SMSS provides a range of aperture magnitudes with a pixel scale four times better than that of ATLAS. We found the PSF magnitudes provide the best correlation with $\qvar$ for non-variable sources. The SMSS observations were made within the period of the light curves.

\begin{figure*}
    \centering
    \includegraphics[width=\linewidth]{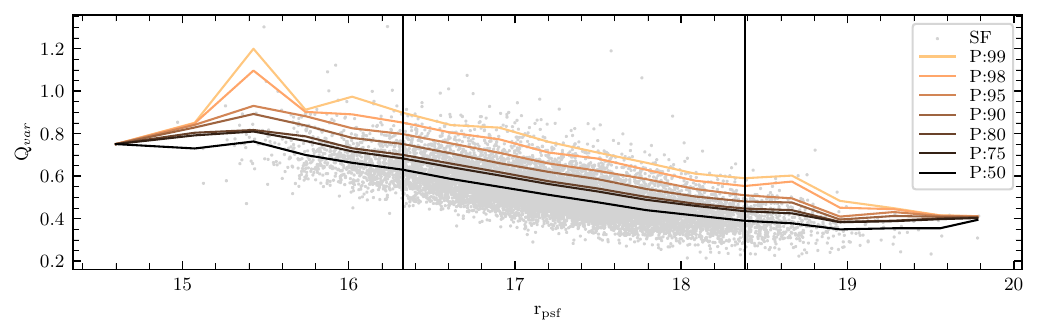}
    \caption{Non-variable objects: effect of source brightness, $r_\mathrm{psf}$, on $\qvar$. The lines mark the percentile contours of SF sources in each 0.3 mag $r_\text{psf}$ bin. The region enclosed by the vertical black lines represents the $r_\text{psf}$ range that is most complete, and we see larger noise in the percentile lines outside this region.}
    \label{fig:percentiles}
\end{figure*}

\subsection{Instrumental noise and transient events}
In order to calibrate for instrumental noise in the light curve measurements, we require a reference sample of sources that do not show any physical variability on the timescales of the light curve period. In this context, the 6dFGS catalogue contains a sample of star-forming (SF) galaxies, with expected non-variable light curves. Since we expect the light curves of type 2 AGN to show a lack of variability similar to those of SF sources, we use the distribution of $\qvar$ of SF sources as a reference for non-variable AGN. As a result, we expect a sensible threshold to be near the upper limit of $r_\text{psf}$-corrected $\qvar$ distribution of SF sources. 
 
For a truly non-variable reference sample, we first need to remove SF galaxies with transient events that would inflate the inferred noise level and thus raise the variability threshold. Separately, we also search for transient events among type 2 AGN to prevent us from finding them later as CLAGN candidates. Transient events, such as supernovae (SNe) and Tidal Disruption Events (TDE), result in a quick rise, followed by a power-law decline in emitted light over time. TDEs are observed in the nucleus of galaxies, as a black hole tears apart an object \citep{gezari_tde_2021}. SNe, in contrast, can be observed in any part of a galaxy, and since we request the centres of the ATLAS photometric aperture onto the nucleus itself, flux measured from an SN at the edge of the galaxy will be underestimated. In both cases, the transient events significantly increase $\qvar$. To identify transients, we visually inspect 305 light curves while descending the highest $\qvar$ values, 39 of which were type 2 AGN. These are the light curves at the upper tail of the $\qvar$ distribution. Our search resulted in 63 events, only five of which were from the type 2 AGN sub-sample. These events are removed from the SF and AGN samples. To prevent any cases of false event assignment, we look for only the most obvious signs of event flares.

\begin{table}
\centering
\caption{Light curve count of 6dFAGN sample used to search for CLAGN after cleaning for undesirable spectra and light curves. The AGN type classification ratios from \citet{winkler_1992} are listed for the BEL AGN.}
\label{tab:agn_count}
\begin{tabular}{ccc}
\hline
6dFAGN Type  & $R=\log\frac{\mathrm{H}\beta\text{-broad}}{\mathrm{[OIII]}}$ & Count \\ \hline
1            & $\phantom{-}0.70 \leq R \phantom{-0.48<}$ & 28  \\
1.2          & $\phantom{-}0.30 \leq R<\phantom{0}0.70$ & 99  \\
1.5          & $-0.48\leq R<\phantom{0}0.30$ & 180 \\
1.8          & $\phantom{000.48<}R<-0.48$ & 14  \\
1.9/2        & - & 1988 \\ \hline
Total        & - & 2309  \\ \hline
Star Forming & - & 12154 \\ \hline
\end{tabular}
\end{table}

\subsection{Variability threshold}
With a defined variability measure, we introduce a threshold that classifies light curves as variable or non-variable. We depict $\qvar$ as a function of $r_\text{psf}$ for SF sources after removing transient events, as illustrated in Figure \ref{fig:percentiles}. The percentile contour lines exhibit linear behaviour, remaining nearly parallel within the most complete magnitude range. However, as the percentiles increase, the contour lines become progressively more erratic due to small number statistics at the high $\qvar$ tail. Linear fits are applied to the contour lines with consistent slopes, while the intercept can be adjusted to derive a threshold, denoted as $\tsf$. This threshold corresponds to the fraction of SF sources positioned below the linear fit line. The linear fit can then serve as a binary classification threshold, where sources located above the threshold are classified as variable, while those below are categorised as non-variable.

The challenge now is to choose an appropriate slope and $\tsf$ for a classification of variable sources that is as pure and complete as possible. We choose the 90th percentile contour line of the SF sources to establish the gradient. This choice optimises proximity to the upper boundary of the magnitude-corrected $\qvar$ distribution while preserving the robustness of the fitted slope. Noise tends to escalate at higher percentiles due to the diminished sample size at the upper boundary. Consequently, the gradient of our chosen cutoff line is set at $-0.134$, mirroring the gradient from the linear fit to the 90th percentile contour line. We then iterate through different intercepts for the cut-off line and compute the completeness and purity of the sources marked as variable.

We define completeness as the fraction of truly variable objects that sit above the variability threshold. Our initial reference list for completeness is constructed by compiling all AGN featuring 6dFAGN types 1-1.5 sourced from the 6dFAGN catalogue, as these are expected to have significant stochastic variability in their light curves \citep{peterson_wanders_1998}. We then clean this list for any known CLAGN with spectroscopic confirmation over the period of the light curves. That is, we remove known turn-off CLAGN and add known turn-on CLAGN from \citet{hon_wolf_2022_skymapper_colors}. The reference list has 248 sources that are expected to show variability. Unknown turn-off CLAGN, if any, are still included in the reference list. Our completeness fraction is therefore marginally underestimated.

To assess the purity of sources marked as variable, we revert to the assumption that none of the SF sources exhibit variability in their light curves. The fraction of SF sources above the threshold line, complementary to $\tsf$, represents the fraction of false positive variable light curves. We exploit this information to compute a contamination rate for the AGN marked as variable. The total number of genuinely non-varying AGN (encompassing all AGN sub-types), denoted as $N_\mathrm{non-var, AGN}$, can be expressed as
\begin{equation}
    N_\mathrm{non-var, AGN} = \frac{1}{\tsf} n_\mathrm{non-var, AGN}.
\end{equation}
Here, $n_\mathrm{non-var, AGN}$ signifies the number of non-variable AGN situated below the threshold line. The fraction of $N_\mathrm{non-var, AGN}$ positioned above the line amounts to ($1-\tsf$). The contamination rate of AGN marked as variable, $C_\mathrm{AGN}$,  is identified as the number of truly non-variable AGN positioned above the threshold line divided by the total number of AGN positioned above this line: 
\begin{equation}
    C_\mathrm{AGN} = \left( 1-\tsf \right) \frac{ N_\mathrm{non-var, AGN}}{n_\mathrm{var, AGN}}.
\end{equation}
Here, $n_\mathrm{var, AGN}$ represents the number of AGN above the threshold line. The purity is the complement of the contamination rate:
\begin{equation}
    \text{Purity} = 1-C_\mathrm{AGN}
\end{equation}

\begin{figure}
    \centering
    \includegraphics[width=\linewidth]{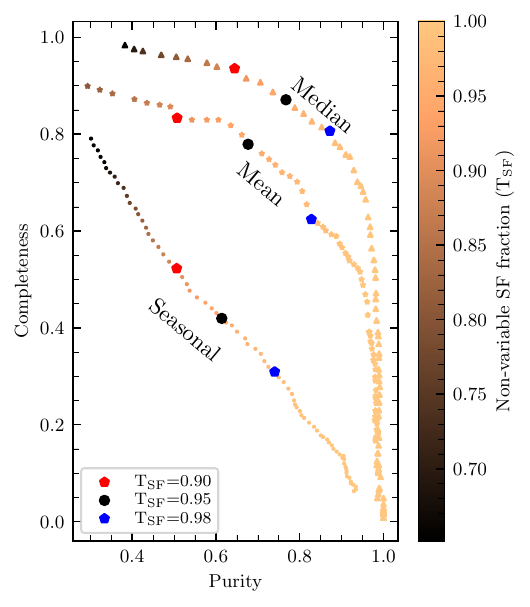}
    \caption{The trade-off between completeness and purity with respect to $T_{SF}$. The three curves are (1) Median and the IQR error light curve stacking used in this work, (2) Mean and standard error light curve stacking which yields a slightly worse variability separation as the mean is sensitive to outliers within the stack, and (3) Seasonal window analysis with median stacking of the light curve which does not provide a satisfactory threshold of variability separation. The three marked points outline a reasonable region for both completeness and purity. We choose $\tsf = 0.95$ as our threshold for the median stacked light curves.}
    \label{fig:completeness}
\end{figure}

We plot the completeness and purity for different $\tsf$ in Figure \ref{fig:completeness} and mark three threshold values where the completeness and purity are suitable. For this work, we choose the threshold line, $\tsf=0.95$, which corresponds to the equation
\begin{equation}
    \qvar = -0.134 \cdot r_\text{psf} + 2.973 \label{eq:t_line},
\end{equation}
with a completeness of 87\% and purity of 77\%. We lean towards higher completeness rather than purity as we can visually check the candidate light curves before observing if the contaminants are numerous.

An expectation of the variability threshold classification is that it preferentially detects truly variable sources from a sample, as it uses a non-variable sample as a reference. However, weak variability that is below the threshold of the instrumental noise is not detected. Sources with variability in a limited period of the light curve would be assigned a high $\qvar$ value, and thus be identified as a variable source. As a consequence, this method is more likely to include high completeness of turn-on CLAGN within the candidates. Conversely, the identification of turn-off CLAGN is more challenging, particularly if the variability state changes within the observed light curve period.

To identify turn-off CLAGN events within the observed light curve period, we attempt to calculate the $\qvar$ metric for each season in the light curves, following a methodology similar to our primary approach. However, the significant overlap between variable and non-variable seasons posed challenges in establishing a robust measure for the seasonal variability $\tsf$. The resulting completeness and purity values failed to yield a reliable separation between variable and non-variable lightcurves (see Figure \ref{fig:completeness}). The sub-period analysis such as the seasonal approach is sensitive to the splitting window length. By ignoring variability over longer periods, smaller window lengths reduce the intrinsic variability signal. Our primary focus is to identify turn-on CLAGN, as other works in the literature mostly start with a BEL AGN search sample and tend to find turn-off CLAGN. Additionally, we have recent spectra for the majority of the variable BEL AGN sample which can be used for CLAGN identification. Therefore, we use the complete light curve period in our analysis. The additional recent spectra are used to check the reliability of the light curve CLAGN candidate selection.

\subsection{Results of variability classification}
\begin{figure*}
    \centering
    \includegraphics[width=\linewidth]{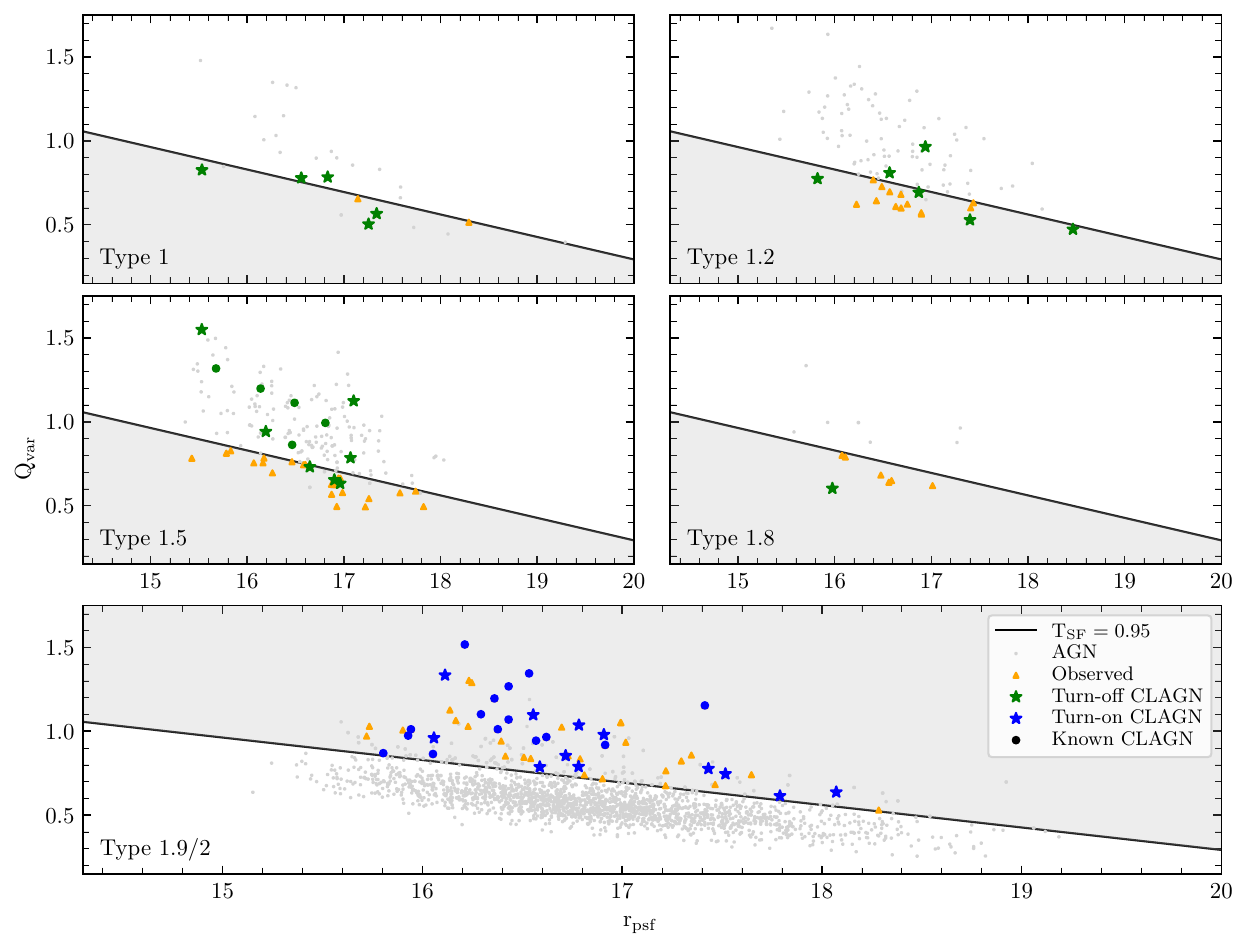}
    \caption{Variability distribution of Seyfert sub-types overlaid over the threshold line corresponding to $\tsf = 0.95$. The grey points depict the AGN distribution. AGN points above the line are classified as variable sources, and points below the line are classified as non-variable sources. The CLAGN candidates are deviations from the expected variability of corresponding sub-types and lie in the shaded region. The triangle markers refer to CLAGN candidates observed with WiFeS which have recent spectral AGN type classification. Turn-off and turn-on CLAGN are denoted by green and blue stars, respectively. We include known CLAGN from \citet{hon_wolf_2022_skymapper_colors} which are within our search sample. Turn-off CLAGN above the threshold line are CLAGN that were identified from existing WiFeS spectra. These CLAGN have turned off during the light curve period, and are above the threshold line due to their variability before the CL event.}
    \label{fig:cut_off_panels}
\end{figure*}

A summary of the variability of AGN sub-types is presented in Figure \ref{fig:cut_off_panels}. Types 1-1.5 show mostly variable behaviour and their $\qvar$ distribution is similar, suggesting variability amplitudes are similar between the three sub-types. The sample size of type 1.8 AGN in our sample does not yield conclusive evidence of whether the expected behaviour of these AGN is variable or non-variable. However, since some broad \hb\ is distinguishable in 1.8s, we choose to select the sources below the threshold as CLAGN candidates. We present the classification of variable and non-variable light curves based on $\tsf=0.95$ in Table \ref{tab:var_result}. Turn-off CLAGN candidates are type 1-1.8 AGN with non-variable light curves and turn-on CLAGN candidates are type 1.9/2 AGN with variable light curves. We identify 201 CLAGN candidates. Based on our completeness estimate, we expect 13\% (40/307) of truly variable type 1-1.5 AGN to be falsely identified as non-variable, so we expect only 18 CLAGN among types 1-1.5 as our completeness only considers these sub-types of variable AGN and it is not possible to provide a similar estimate for type 1.8 AGN. For type 1.9/2 AGN, we expect 5\% ($1-\tsf$; 99/1988) false positive CLAGN candidates.
\begin{table}
\centering
\caption{Variability by AGN Type. CLAGN candidates are marked in bold.}
\label{tab:var_result}
\begin{tabular}{cccc}
\hline
6dFAGN & Variable & Non-variable & CLAGN Candidate \\
Type & AGN & AGN & Fraction \\ \hline
1         & 19              & \textbf{9}        & 0.321    \\
1.2       & 80              & \textbf{19}       & 0.192    \\
1.5       & 150             & \textbf{30}       & 0.167    \\
1.8       & 7               & \textbf{7}        & 0.500    \\ \hline
1.9/2     & \textbf{136}    & 1852              & 0.068    \\ \hline
\end{tabular}
\end{table}
\section{Spectroscopic Follow-Up and Analysis}\label{sec:spec-follow-up}

\begin{table*}
\centering
\caption{New CLAGN identified in this work. Redshift is from 6dFGS, r$_\text{psf}$ from SMSS DR3, 6dFGS and WiFeS AGN Type and line ratio from emission line fitting with BADASS3. Type 2 implies no broad H$\alpha$ emission seen by visual examination. 2$\sigma$ upper limits are reported for the \hb/[OIII] ratio at the epoch where broad \hb\ line is not observed. An asterisk ($^*$) denotes CLAGN that were not identified as candidates. For two turn-on CLAGN, g2051395-175306 and g2258019-021945, we visually found a lack of broad lines, although our conservative upper limits for the 6dFGS \hb/[OIII] ratio does not prove a change in AGN type (see spectra in Figure \ref{fig:new_on_cls_2}).}
\label{tab:new_clagn_lc}
\begin{tabular}{lrccccccccc}
\hline & & & & \multicolumn{3}{c}{\textbf{6dFGS}} & & \multicolumn{3}{c}{\textbf{WiFeS}} \\ \cline{5-7} \cline{9-11} 
\begin{tabular}[c]{@{}c@{}}6dFGS \\ Name\end{tabular} & \begin{tabular}[c]{@{}c@{}}6dFGS \\ Spec ID\end{tabular} & z & r$_\text{psf}$ & Epoch (MJD) & Type & $\log \frac{\text{H}\beta}{\text{[O{\tiny III}]}}$ && Epoch (MJD) & Type & $\log \frac{\text{H}\beta}{\text{[O{\tiny III}]}}$ \\ \hline
\multicolumn{11}{c}{\textbf{Turn-on CLAGN}} \\ \hline
g0144586-023159 & 11311  & 0.0958 & $16.606\pm0.044$ & 52610 & 1.9 & $<-0.17$ && 60171 & 1.5 & $\phantom{-}0.06^{+0.12}_{-0.14}$ \\[5pt]
g0209537-135321 & 13500  & 0.0726 & $16.784\pm0.064$ & 53248 & 1.9 & $<-0.09$ && 59905 & 1.5 & $\phantom{-}0.16^{+0.28}_{-0.39}$ \\[5pt]
g0618343-413916 & 35723  & 0.0825 & $16.735\pm0.081$ & 53033 & 1.9 & $<-0.39$ && 60021 & 1.5 & $-0.12^{+0.07}_{-0.08}$           \\[5pt]
g1007504-090445 & 47853  & 0.0669 & $16.947\pm0.082$ & 52759 & 1.9 & $<-0.03$ && 59725 & 1.5 & $\phantom{-}0.11^{+0.08}_{-0.08}$ \\[5pt]
g1038209-100660 & 50712  & 0.0289 & $16.203\pm0.096$ & 53407 & 1.9 & $<-0.14$ && 59732 & 1.2 & $\phantom{-}0.34^{+0.03}_{-0.03}$ \\[5pt]
g1415440-220050 & 71677  & 0.0494 & $17.432\pm0.023$ & 52827 & 1.9 & $<-0.02$ && 59726 & 1.5 & $\phantom{-}0.08^{+0.06}_{-0.07}$ \\[5pt]
g1423582-050009 & 73242  & 0.0558 & $17.034\pm0.069$ & 53470 & 1.9 & $<-0.76$ && 59726 & 1.5 & $-0.30^{+0.05}_{-0.06}$           \\[5pt]
g1548163-175935 & 80509  & 0.0893 & $17.577\pm0.065$ & 52798 & 1.9 & $<-0.28$ && 60166 & 1.2 & $\phantom{-}0.50^{+0.07}_{-0.07}$ \\[5pt]
g2051395-175306 & 97930  & 0.0877 & $18.164\pm0.049$ & 52884 & 2.0 & $<-0.09$ && 60083 & 1.5 & $-0.11^{+0.15}_{-0.21}$           \\[5pt]
g2109165-432059 & 100263 & 0.0940 & $17.790\pm0.034$ & 52880 & 1.9 & $<-0.32$ && 60167 & 1.2 & $\phantom{-}0.58^{+0.08}_{-0.08}$ \\[5pt]
g2202067-053946 & 105367 & 0.0271 & $16.084\pm0.050$ & 53262 & 1.9 & $<-0.12$ && 60093 & 1.5 & $-0.06^{+0.07}_{-0.07}$           \\[5pt]
g2258019-021945 & 132260 & 0.0800 & $16.583\pm0.058$ & 53238 & 1.9 & $<-0.05$ && 60176 & 1.5 & $-0.05^{+0.11}_{-0.11}$           \\ \hline
\multicolumn{11}{c}{\textbf{Turn-off CLAGN}} \\ \hline
g0010200-061706$^*$ & 501    & 0.0820 & $16.546\pm0.035$ & 52966 & 1.0 & $\phantom{-}0.83^{+0.11}_{-0.11}$ && 60225 & 2.0 & $<-0.03$ \\[5pt]
g0101244-030840$^*$ & 121947 & 0.0701 & $16.568\pm0.080$ & 53265 & 1.2 & $\phantom{-}0.46^{+0.07}_{-0.07}$ && 60223 & 2.0 & $<-0.66$ \\[5pt]
g0200374-133040     & 12516  & 0.0530 & $16.903\pm0.145$ & 52991 & 1.5 & $\phantom{-}0.29^{+0.09}_{-0.11}$ && 60171 & 1.9 & $<-0.47$ \\[5pt]
g0231241-250511     & 14201  & 0.0522 & $16.532\pm0.049$ & 52529 & 1.5 & $-0.34^{+0.16}_{-0.28}$           && 60178 & 1.9 & $<-0.60$ \\[5pt]
g0259401-184736$^*$ & 17861  & 0.0697 & $17.102\pm0.043$ & 52961 & 1.5 & $\phantom{-}0.03^{+0.07}_{-0.08}$ && 60181 & 1.9 & $<-0.65$ \\[5pt]
g0645395-402129     & 126074 & 0.0352 & $16.939\pm0.062$ & 53080 & 1.2 & $\phantom{-}0.48^{+0.11}_{-0.12}$ && 59668 & 1.9 & $<-0.09$ \\[5pt]
g0913233-255925     & 43999  & 0.0546 & $17.206\pm0.045$ & 52642 & 1.0 & $\phantom{-}0.97^{+0.52}_{-0.83}$ && 59731 & 2.0 & $<-0.32$ \\[5pt]
g0936221-113436$^*$ & 118704 & 0.0913 & $16.900\pm0.036$ & 52763 & 1.2 & $\phantom{-}0.61^{+0.05}_{-0.05}$ && 60019 & 1.9 & $<-0.23$ \\[5pt]
g1254564-265702$^*$ & 64181  & 0.0592 & $15.515\pm0.024$ & 52822 & 1.5 & $-0.08^{+0.05}_{-0.04}$           && 59725 & 1.9 & $<-0.50$ \\[5pt]
g1323403-012749$^*$ & 66221  & 0.0768 & $17.031\pm0.057$ & 53471 & 1.5 & $\phantom{-}0.15^{+0.07}_{-0.08}$ && 60081 & 1.9 & $<-0.44$ \\[5pt]
g1556253-202829$^*$ & 80577  & 0.0618 & $16.859\pm0.030$ & 52798 & 1.0 & $\phantom{-}1.55^{+0.44}_{-0.56}$ && 59725 & 1.9 & $<-0.03$ \\[5pt]
g1619441-132616     & 81958  & 0.0788 & $18.747\pm0.137$ & 53112 & 1.2 & $\phantom{-}0.65^{+0.21}_{-0.25}$ && 60165 & 1.9 & $<-0.43$ \\[5pt]
g1630074-001136     & 82577  & 0.0466 & $17.024\pm0.087$ & 53138 & 1.5 & $\phantom{-}0.02^{+0.07}_{-0.07}$ && 59725 & 1.9 & $<-0.49$ \\[5pt]
g2139427-290315     & 102100 & 0.0729 & $17.406\pm0.079$ & 52499 & 1.0 & $\phantom{-}0.71^{+0.10}_{-0.11}$ && 60166 & 2.0 & $<-0.45$ \\[5pt]
g2142147-162948$^*$ & 103713 & 0.0515 & $16.194\pm0.034$ & 53237 & 1.5 & $\phantom{-}0.21^{+0.13}_{-0.14}$ && 60168 & 1.9 & $<-0.44$ \\[5pt]
g2154290-282145     & 105503 & 0.0326 & $15.533\pm0.056$ & 52530 & 1.0 & $\phantom{-}0.95^{+0.16}_{-0.16}$ && 60168 & 1.9 & $<\phantom{-}0.09$ \\[5pt]
g2156566-113931     & 104501 & 0.0281 & $15.977\pm0.041$ & 53261 & 1.8 & $-0.71^{+0.13}_{-0.15}$           && 60164 & 2.0 & $<-1.28$ \\[5pt]
g2205311-373711     & 106162 & 0.0565 & $17.316\pm0.126$ & 52470 & 1.2 & $\phantom{-}0.44^{+0.11}_{-0.12}$ && 60083 & 2.0 & $<-0.13$ \\[5pt]
g2253107-040849     & 132277 & 0.0253 & $15.900\pm0.045$ & 53238 & 1.2 & $\phantom{-}0.64^{+0.07}_{-0.07}$ && 59026 & 1.9 & $<-0.10$ \\\hline
\end{tabular}
\end{table*}

\subsection{Spectrograph and processing of spectral data}

We conduct follow-up observations of our CLAGN candidates using emission-line spectroscopy to confirm potential changes in AGN sub-types. All observations were carried out at the Australian National University (ANU) 2.3m telescope situated at the Siding Spring Observatory. The observations utilised the Wide Field Spectrograph \citep[WiFeS;][]{dopita_hart_2007,dopita_rhee_2010}, an integral field instrument with a field-of-view of $38\times25$ arcsec with 1 arcsec$^2$ spaxels. We used the B3000 and R3000 diffraction gratings, resulting in a total wavelength range of $3200 - 9800$\AA. The integral field data cubes were reduced using PyWiFeS \citep{childress_vogt_2014_pywifes}. 

WiFeS provides much better spatial and spectral resolution compared to the 6.7" diameter aperture from 6dFGS with R$\sim$1000. Consequently, 6dFGS includes a higher fraction of stellar light compared to a single WiFeS spaxel at the nucleus at $z<0.1$. Even for narrow emission lines, the line ratios will depend on the parts of the galaxy included in the spectrum \citep[see the mixing sequence in][]{davies_groves_2016}. To probe the same physical footprint as the 6dFGS observations, we sum light from a 37 spaxel circular aperture from the WiFeS data cube assuming the 6dFGS aperture is centered on the nucleus of the galaxy. A comparison between $\log$(\hb/\oiii) ratios of WiFeS pixel scale and 6dFGS aperture is presented in Appendix \ref{sec:cen_vs_wif}, showing that the ratio obtained from the central spaxel results in a 12\% increase in the ratio compared to the 6.7" aperture. Comparing positions with SMSS shows less than 1" offset in the reported 6dFGS fibre positions for a majority of the sources. Additionally, WiFeS spectra are re-binned to a lower resolution to align with the resolution of the 6dFGS spectra, using the Python \texttt{spectres} package \citep{carnall_2017_spectres}.

The 6dFGS spectra are not flux calibrated. Since we expect minimal to no variations in emission from the narrow line region and collective stellar emission across the galaxy over timescales as short as a decade, we have chosen to normalise the 6dFGS spectra to the flux-calibrated WiFeS spectra using the $\lambda5007$\AA\oiii\ emission line. However, since CLAGN identification relies on the \hb/\oiii\ line ratio, and the invariance of the \oiii\ line on short timescales is established \citep{groningen_wanders_1992}, we can mostly disregard the effects of flux normalisation in CLAGN identification. Since 6dFGS spectra are not flux calibrated, the detector sensitivity results in significantly bluer spectra relative to the WiFeS spectra which are calibrated using standard stars. This effect can be ignored as the \oiii\ line is close enough to \hb\ line that the continuum slope variations only change the \hb\ flux marginally. The change in wavelength between \hb\ and \oiii\ is $\sim$3\%, so a slope of 1 gives a 3\% flux change. This is reduced to $<$1\% flux change for a slope of $<$0.33, as is typically seen in this wavelength region in our spectra, which is much smaller than the errors on the \hb/\oiii\ ratio.

\subsection{CLAGN identification}
\begin{figure*}
    \centering
    \includegraphics[width=0.98\linewidth]{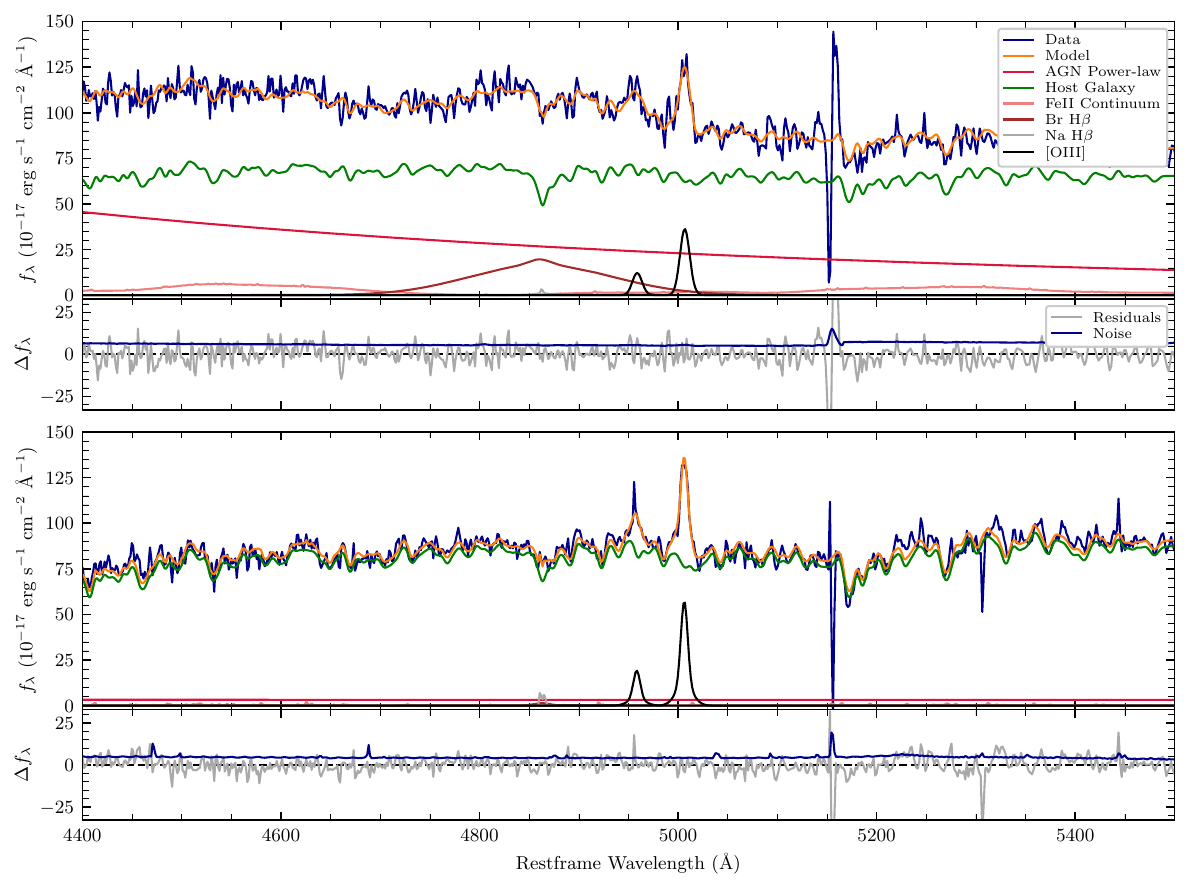}
    \caption{Examples of the best-fit models of spectra, fit around the \hb/[OIII] region with the fitting components marked by different colours. This particular AGN is g0010200-061706, a turn-off CLAGN, changing from type 1 (6dFGS; \textbf{top}) to type 2 (WiFeS; \textbf{bottom}). For clarity, the individual emission line Gaussian components are summed and only the final components are plot. Note that the 6dFGS spectrum has been flux calibrated using the [OIII] peak of the WiFeS spectrum.}
    \label{fig:fit_ex}
\end{figure*}

\begin{figure*}
    \centering
    \includegraphics[width=\linewidth]{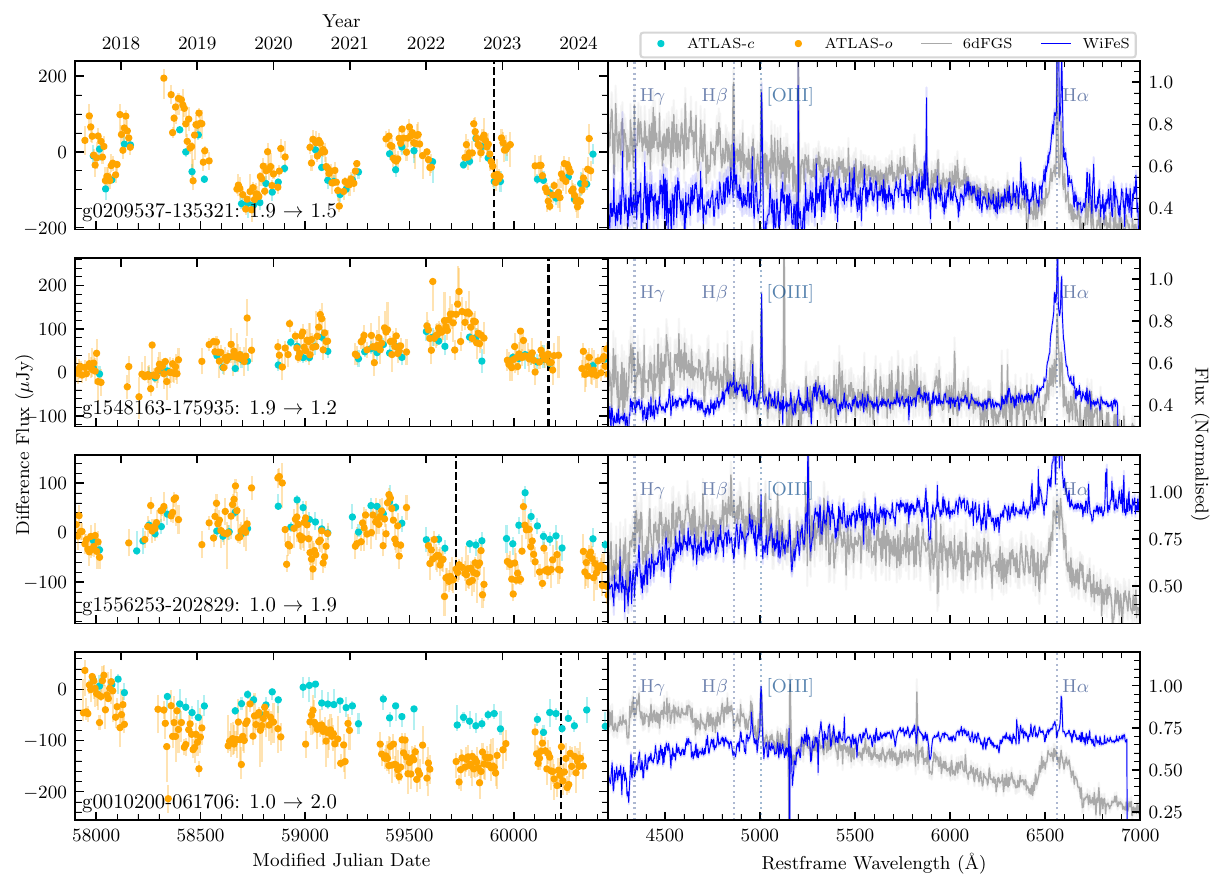}
    \caption{\textbf{Left:} ATLAS light curves for a representative sample of CLAGN, identified by the 6dFGS object name, separated into cyan and orange filters. Error bars are interquartile ranges of measurements within stacking periods. The dashed vertical line marks the epoch of the WiFeS observation. \textbf{Right:} 6dFGS (grey) and WiFeS (blue) spectra of the same CLAGN. The 6dFGS $\rightarrow$ WiFeS epoch spectral sub-type classification is printed on the light curve panel. All spectra are normalised to the $\lambda 5007$\AA{} line. Note that since the 6dFGS spectra are not completely calibrated, their continuum may show features of the detector sensitivity, resulting in bluer spectra compared to WiFeS.}
    \label{fig:rep}
\end{figure*}

We used Bayesian AGN Decomposition Analysis for SDSS Spectra \citep[BADASS3\footnote{https://github.com/remingtonsexton/BADASS3};][]{sexton_matzko_2021_badass3} to fit multiple Gaussian components to emission lines and extract broad \hb\ and $\lambda5007$\AA\oiii\ luminosity from de-reddened spectra. For consistency, we fit both 6dFGS and WiFeS with the same fit parameters using BADASS3. The fitting algorithm was restricted to a restframe wavelength range from 4400\AA\ to 5500\AA. The free parameters for the spectral fitting include an AGN power law, single-stellar population host models, along with a broad and narrow optical FeII continuum. Broad emission lines for AGN are fit using multiple Gaussian components and summed up, resulting in a combined model \citep[for example, see][]{greene_ho_2005,shen_richards_2011,rakshit_stalin_2020}. For the emission lines, we chose two narrow (FWHM $<500$ km s$^{-1}$) Gaussian components each for \hb\ and \oiii\ and an additional three broad (FWHM $>500$ km s$^{-1}$) Gaussian components for \hb. The FWHM constraints here are specifically for each Gaussian component and not for the combined model. Only the broad component is used to obtain the \hb/\oiii\ line ratio and since by definition there is no broad \hb\ component in types 1.9 and 2 AGN, we report 2$\sigma$ upper limits for the ratio. The limits are computed fitting an additional Gaussian component with fixed FWHM obtained from the epoch with broad lines. For this work, we define CLAGN as AGN with appearance (turn-on) or disappearance (turn-off) of broad \hb\ components with combined FWHM $>1200$ km s$^{-1}$ \citep[following][]{hao_strauss_2005,shen_richards_2011}, i.e., AGN with type changes between types \{1, 1.2, 1.5, 1.8\} and \{1.9, 2\} across the 6dFGS and WiFeS epochs. See Figure \ref{fig:fit_ex} for example best-fit models of type 1 and type 2 spectra. We emphasise that we fit only the region around \hb\ using \oiii\ to normalise between the two epochs. On the red end of the spectra, the lack of calibration in the 6dFGS spectra, along with a lack of a strong emission line such as the \oiii\ to normalise the red end, makes it difficult to consistently compare the H$\alpha$ emission. As a consequence, we do not make the distinction between type 1.9 and type 2 AGN in our quantitative comparisons. In Table \ref{tab:new_clagn_lc}, Figure \ref{fig:rep}, and subsequent appendix figures, the reported difference between types 1.9 and 2 is through a visual check on the presence of broad H$\alpha$. While this may be a limited and rather subjective approach, we do not report any quantitative results that distinguish between the type 1.9 and type 2 AGN in this work.

We have recent WiFeS spectra for 327 AGN within our light curve search sample, 106 of which were selected as CLAGN candidates in this work. These spectra date from 2018, but we performed targeted candidate observations between May 2022 and October 2023, and we preferentially observed targets that showed candidate-like variability behaviour towards the recent end of the light curve after a visual check of the variability. Our candidate list includes 17 AGN which had been previously observed as CLAGN candidates selected by an alternative method \citep{hon_wolf_2022_skymapper_colors}, and observations for these were not duplicated. 

Based on our fits to the \hb\ and \oiii\ lines, we find 31 new CLAGN with light curves in ATLAS, 23 of which were selected as candidates. Observed CLAGN candidates and CLAGN are presented in Figure \ref{fig:cut_off_panels} along with known CLAGN in the literature within our light curve search sample. We also find 13 turn-off CLAGN which were `missed' (not candidates). The `missed' turn-off CLAGN have turned-off during the light curve period, with the period of initial variability increasing the measured $\qvar$. Five of these CLAGN were identified in \citet{hon_wolf_2022_skymapper_colors}, with the remaining eight being newly identified CLAGN in this work. None of the observed non-variable type 1.9/2 sources show broad \hb\ lines in their recent spectra. Including the 15 known turn-on CLAGN from \citet{hon_wolf_2022_skymapper_colors} in our candidates, 36\% (38/106) of light curve-selected CLAGN candidates are CLAGN with spectroscopic confirmation. Considering only the turn-on CLAGN, which was our primary focus, our CLAGN candidate success rate increases to 51\% (27/53). Here, the success rate is the fraction of spectroscopic CLAGN among observed CLAGN candidates, where we have observed all likely CLAGN candidates after a visual check of the light curve. A complete list of new CLAGN is presented in Tables \ref{tab:new_clagn_lc}. The light curves and spectra of a representative sample of four CLAGN are shown in Figure \ref{fig:rep} and one special case is shown in Figure \ref{fig:g1254564-265702}. All remaining CLAGN discovered in this work are presented in Appendix \ref{sec:all_cl_plots}.

\section{Discussion}\label{sec:discussion}

\subsection{CLAGN in Southern sky}
In this work, we identify CLAGN solely by the appearance or disappearance of the broad \hb\ line over the two spectral epochs. This excludes variations in already existent broad \hb\ line, corresponding to changes within the sub-types 1-1.8. Similar to all CLAGN identification studies, we miss CLAGN with double CL events occurring between the spectral epochs, which change and change back before a spectrum is taken. Any information on AGN behaviour is lost between the 6dFGS epoch (2001-2009) and before the ATLAS light curve era (2016). This timescale is longer than some of the known double CL events \citep[for example, see][]{tohline_osterbrock_1976,ricci_kara_2020,zeltyn_trakhtenbrot_2022}. However, most of these double CL events turned off and then on, as these studies, and many other current systematic CLAGN searches, start with AGN samples already containing broad \hb\ lines. Here, we start with a type 2 sample, with no information on whether they are "true" type 2 or AGN with hidden BLR. Therefore, we expect to miss rapid turn-on $\rightarrow$ turn-off double CL events. However, these rapid flares could either be Tidal Disruption Events (TDEs), or short-lived increases in accretion resulting in flares which may be a new class of AGN events different from CLAGN \citep{trakhtenbrot_arcavi_2019}.

We are also limited by the spectral quality in identifying weak broad lines. The lack of detection of these broad lines would classify these objects as type 1.9/2. Our ability to find CLAGN is also limited to AGN with unobscured lines-of-sight to the BLR. Any AGN with orientation-induced obscuration would persistently be classified as type 2, regardless of the activity level. The 32 new CLAGN we have identified within our search sample are thus an incomplete snapshot of the CL events that have changed types between the 6dFGS and ATLAS periods. While we may find new CLAGN with another two epoch study with a different timescale, between this work and \citet{hon_wolf_2022_skymapper_colors}, we expect to have found a majority of recent CL events at $z<0.1$ and $-45\degree<\delta<0\degree$ from the 6dFGS parent sample. Any new optical CLAGN within this region are only expected from the remaining 69 BELAGN without WiFeS data, where our light curve method is not very effective as we cannot find turn-off sources from light curves with periods of both variability and non-variability.

\subsection{CL event rates}
Our analysis revealed 12 new turn-on CLAGN, on top of 15 already known turn-on CLAGN within our light curve sample. This establishes a lower limit of 1.4\% (27/1988) turn-on rate over $\sim$15 years. This conservative rate does not account for turn-on CL events in orientation-induced obscured AGN. A more reliable metric for CL events can be obtained from turn-off CLAGN, where the presence of broad lines at any point in time indicates an unobscured line-of-sight to the BLR. With the 19 new turn-off CLAGN we find, and in addition to 5 known turn-off CLAGN in our sample, we estimate a turn-off CLAGN rate of at least 9.6\% (24/251) over $\sim$15 years, corresponding to 0.6\% yr$^{-1}$. Here we computed the fraction from just the 251 observed BEL AGN from 6dFGS, since $>$75\% of the BEL AGN were observed and the right ascension of the object was the dominant limiting factor for the unobserved sources, thus minimising any selection effects. For the turn-on candidates, only the most likely candidates were observed after visual examination because it was not feasible to observe all 1,988 AGN.

Since we have a good sampling of the observed BEL AGN, we can use it to estimate the completeness of our $\tsf$ threshold on our $\qvar$ for separating variable light curves. Excluding CLAGN, we find 81\% of the variable BEL AGN lie above the $\tsf=0.95$ line, while our blind estimate using just the light curves based on 6dFAGN-epoch sub-type classification predicted 87\% completeness. Using the new completeness, only 81\% of the turn-on CLAGN have been found from our type 1.9 and 2 AGN. Correcting for the variability completeness, we estimate 33 turn-on CLAGN, resulting in a corrected turn-on CL rate of 1.7\% per 15 years. 

Our CL event rates are consistent with \citet{lopez_martinez_2022}, who use a similar light curve variability approach to find turn-on CLAGN. They identify 4 CLAGN after observing 6 AGN from a candidate pool of 30 type 2 AGN and predict from their complete sample that 1.8\% of type 2 AGN have appearing BELs over $\sim$15 years. In their following work, \citet{lopez-navas_sanchez-saez_2023} estimate a slightly higher turn-on rate of 3\% over 15 years, by identifying 15 CLAGN from a candidate list of 30, and extrapolating their selection cuts back to their parent sample of $\sim$30,000 AGN. Our turn-on rate is marginally lower due to the completeness of our type 2 AGN parent sample. Our CL event rates are consistent with the reported rate of 3\% over 15 years for both turn-on (35/1092) and turn-off (11/314) CLAGN by \citet{hon_wolf_2022_skymapper_colors}, who use a search sample that overlaps with this work. \citet{runco_cosens_2016} also find a 3\% (3/102) CL event rate with complete BEL disappearance over 3$-$9 years using a sample at a similar redshift to this work. \citet{temple_ricci_2023_clagn} find a similar combined CL rate of 0.7\%$-$6.2\% over 10$-$25 years (8/749 to 21/412 including Poisson statistical uncertainties). They report a range due to a heterogeneous parent search sample with different selection functions. A recent study by \citet{zeltyn_trakhtenbrot_2024} reports a lower limit on the CL rate of 1.2\% over 20 years for $z<1$, while also presenting the luminosity dependence of CL rates ranging approximately from 10\% to 1\% for $\mathrm{L}_\mathrm{bol} \sim 10^{43.8-45.4}$ (see their Figure 4). For the median luminosity $\mathrm{L}_\mathrm{bol} \sim 10^{44.4}$ erg s$^{-1}$ for our BEL AGN sample, this corresponds to a predicted CL rate of $\sim$6\% over 20 years from the $z<0.5$ dataset of \citet{zeltyn_trakhtenbrot_2024}.

We note that the number of turn-on (27) and turn-off (24) events is the same within Poisson noise. This is expected, if we assume that (i) our AGN sample is complete in the survey volume of 6dFGS over the redshift interval of $z<0.1$ (AGN in dwarf galaxies would be missing from the sample), and (ii) the timescales for the type change are shorter than the separation of the observing epochs, and thus both types of events would have the same chance to be detected.

The process of changing AGN accretion states is also related to the population of unobscured type 1.9/2 AGN, as every turn-off event produces an AGN of the latter type, until it turns back on. Unfortunately, we cannot derive meaningful constraints on unobscured type 1.9/2 AGN from our observations. If we assumed that AGN spent the same fraction of time in on-state and off-state, then we ought to have as many unobscured type 1.9/2 AGN as type 1-1.8 objects, which would imply an unobscured fraction of $\sim$16\% (321/1988) among all type 1.9/2 AGN. While this number is broadly consistent with other estimates of the unobscured type 1.9/2 AGN fraction in the literature \citep[e.g.][]{panessa_bassani_2002,garcet_gandhi_2007,brightman_nandra_2008}, our estimate depends entirely on the unconstrained time fraction spent by AGN in different states. A more detailed study of the X-ray properties of our type 1.9/2 sample is in progress.

Whether all AGN are CLAGN or only a static subset of AGN produce CL events is currently unconstrained. With a 0.6\% yr$^{-1}$ CL event rate, identifying all CLAGN within a complete sample would take more than 150 years of continuous monitoring. If the CL event rate remains constant, and the fraction of known CLAGN rises steadily while correcting for the turn-on CL rate, we expect that all AGN can occasionally exhibit CL behaviour. If only a subset of the AGN show CL behaviour, we would expect the CLAGN fraction to plateau before reaching a hundred per cent and the CL event rate can be used to estimate the timescale between CL events within a single CLAGN.

\subsection{TDE induced turn-off event}
\begin{figure*}
    \centering
    \includegraphics[width=\linewidth]{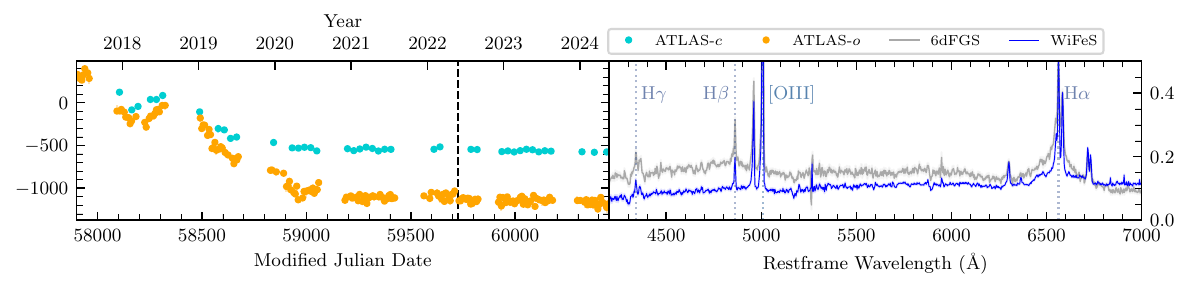}
    \caption{6dFGS object g1254564-265702, similar to Figure \ref{fig:rep}. A possible TDE-induced decline in flux between variable and non-variable phases of the light curve is visible between MJD 58500 and 59000. The event possibly peaked when the object was behind the Sun.}
    \label{fig:g1254564-265702}
\end{figure*}

The power-law decline of TDEs, which are sometimes argued to cause type changes in AGN \citep[see][]{merloni_dwelly_2015,ricci_kara_2020}, is only observed in one object in our sample, where a decline in flux possibly coincides with the CL event (g1254564-265702; Figure \ref{fig:g1254564-265702}). The 6dFAGN spectrum shows type 1.5 features, but the recent WiFeS spectrum shows broad H$\alpha$ and narrow \hb\ lines, indicating type 1.9. The light curve shows variability before mid-2018, and a transient event \citep[Transient 2019hsf;][]{delgado_2019_2019hsf} which might have a peak when the object was behind the Sun. There is no variability after the decline corresponding to the lack of broad lines in the WiFeS spectra. The variability before 2019 suggests the AGN was somewhere between type 1.0 and 1.8, but there is no spectroscopic evidence.

Differentiating TDEs and already variable AGN in high activity states is not straightforward \citep{auchettl_ramirez_2018,trakhtenbrot_arcavi_2019}. Particularly when a tidally disrupted star may enhance or even stop the accretion by interacting with the existing accretion disc \citep{li_ho_2022,chan_piran_2019}. In the case of g1254564-265702, the rise in flux starting around MJD 58250 suggests a year-long flare within a longer period of continuous dimming of the AGN. The extended period of weak variability after the dimming corresponds to type-2 behaviour seen in most other light curves.

\subsection{CLAGN mechanism}
The current studies within the literature find that CL events are either Changing accretion State (CS) or Changing Obscuration (CO). From our analysis of the CL events within the light curve periods, we find the following:
\begin{enumerate}
    \renewcommand{\labelenumi}{(\roman{enumi})}
    \item Occasionally, turn-off events in the light curve are preceded by a prolonged flare similar to the findings in \citet{runnoe_cales_2016}. For example, see light curves of g0101244-030840, g0231241-250511 (Figure \ref{fig:new_off_cls_1}),  g0936221-113436 (Figure \ref{fig:new_off_cls_2}) and g2205311-373711 (Figure \ref{fig:new_off_cls_3}) among the turn-off CLAGN in the Appendix. It has been speculated that this prolonged activity of high emission is due to the rapid accretion of the inner region of the accretion disc. The breaking down of the inner accretion disc is discussed in \citet{ricci_kara_2020} and also shown in simulations \citep[eg.,][]{nealon_price_2015,kaaz_liska_2023}. CL events that result in a type 2 AGN are extreme cases where the accretion disc is wholly disrupted. This model also explains the few observed cases where TDEs cause turn-off CL events. CL events are preferentially observed within low luminosity AGN, and similarly, CL quasars are observed at lower Eddington ratios \citep{macleod_green_2019}, suggesting smaller accretion discs which are relatively easier to disrupt by disturbances around the accretion disc. This model favours the CS phenomenon of CL events, as the prolonged flare in the light curves before a turn-off event would not be visible during a transient obscuration. However, these flares could just be the most recent high state in the AGN stochastic variability itself, and the decline could partially be the shutting down of accretion, or the onset of a passing obscurer.
    \item Turn-on CL events usually show a gradual rise in flux, followed by the expected stochastic variability of BEL AGN. This is a natural follow-up of the inner accretion disc-breaking scenario, as the outer disc fills in and regenerates the complete disc. However, the gradual increase in flux could also be the result of the tail end of a passing dust or gas cloud.
\end{enumerate}

We also notice certain trends in the light curve behaviour during the time of the WiFeS observations. Spectroscopic observations of variable AGN contain broad lines, even if observed during a declining period of the variability. AGN observed during a temporary low emission state do not contain broad \hb\ emission, but do contain broad H$\alpha$ suggesting type 1.9 AGN are a temporary decline in AGN activity, or modest obscuration which mostly affect \hb\ observations. Type 2 AGN are observed during a prolonged period of non-variability suggesting either complete obscuration or a true lack of AGN activity. Examples of these observations are presented in Figure \ref{fig:rep}.
\section{Conclusion}\label{sec:conclusion}

The frequency and timescale of changing-look (CL) events are crucial to understanding AGN evolution on decadal timescales. To better constrain these parameters, we need a large sample of CLAGN. In this work, we combined historic spectra of AGN from the 6dF Galaxy Survey, which is a large, complete, flux-limited hemispheric survey, with recent light curves from NASA-ATLAS to search for CLAGN: historical BEL AGN with recently non-variable light curves as well as historical narrow-line AGN with recently variable light curves are both strong candidates for CLAGN. These candidates were observed with WiFeS to confirm an appearance or disappearance of the broad \hb\ line. For the first time, we search for turn-on CLAGN in a magnitude complete galaxy sample covering one-third of the visible sky, at $z<0.1$ and $-45\degree < \delta < 0\degree$; starting from the 6dF Galaxy Survey, we work with a type 2 AGN sample of unprecedented completeness.

We identified 31 new CLAGN, 23 of which were within our candidate list. After including previously known CLAGN within our candidate list, we report a 51\% success rate for turn-on CLAGN identified using the light curve method. We find lower limits for CL event rates over 15$-$20 years as 1.7\% for turn-on and 9.6\% for turn-off CLAGN, where the turn-on rate is not corrected for type 1.9/2 AGN that are obscured due to orientation. The CLAGN rates we find agree with similar studies within the literature. The number of turn-on (27) and turn-off (24) events is the same within Poisson noise, which is expected assuming that our AGN sample is reasonably complete in the $z<0.1$ search volume.

With a visual examination of the light curves, we show that broad lines are observed during periods of variability, even when the AGN is in a relatively low state, while type 2 AGN appear during extended periods of non-variability. This can be used to estimate AGN states during epochs where only photometric data is available. From the light curves, we find that turn-off events are occasionally marked by a prolonged flare followed by a decline in variability and flux to a low activity state, and turn-on events are a gradual rise in flux and variability. These observations favour the changing-state (CS) flavour of CL events. Since ATLAS covers the entire Northern Sky and coincides with a majority of large-scale CLAGN searches, the effort of examining the light curves of these CLAGN would be worthwhile in understanding AGN light curve behaviour around CL events.

CLAGN at higher redshifts are still an area to explore in the Southern sky as a majority of large-scale multi-epoch surveys target the Northern Sky. Ideal CLAGN identification efforts would utilise high cadence multi-epoch spectroscopic surveys targeting a large sample of known AGN of all sub-types. Upcoming large-scale time domain surveys such as the SDSS-V Black Hole Mapper \citep{SDSS_V_plan} and LSST at the Vera C. Rubin Observatory \citep{LSST_plan} are expected to significantly develop our understanding of short-term AGN evolution and CL events.

\section*{Acknowledgements}
The authors thank the anonymous reviewer for helpful comments that improved the quality of the paper.
NA was supported by Australian Government Research Training Program (RTP) Scholarship.
CAO was supported by the Australian Research Council (ARC) through Discovery Project DP190100252.
This paper is based on observations made with the Australian National University 2.3m Telescope at Siding Springs Observatory. We thank the WiFeS observers for efforts in acquiring the spectra for this paper. The observers include the authors in this paper, as well as Katie Auchettl, Patrick Tisserand and Harrison Abbot.
This work has made use of data from the Asteroid Terrestrial-impact Last Alert System (ATLAS) project. The ATLAS project is primarily funded to search for near earth asteroids through NASA grants NN12AR55G, 80NSSC18K0284, and 80NSSC18K1575; byproducts of the NEO search include images and catalogs from the survey area. This work was partially funded by Kepler/K2 grant J1944/80NSSC19K0112 and HST GO-15889, and STFC grants ST/T000198/1 and ST/S006109/1. The ATLAS science products have been made possible through the contributions of the University of Hawaii Institute for Astronomy, the Queen’s University Belfast, the Space Telescope Science Institute, the South African Astronomical Observatory, and The Millennium Institute of Astrophysics (MAS), Chile.
The national facility capability for SkyMapper has been funded through ARC LIEF grant LE130100104 from the Australian Research Council, awarded to the University of Sydney, the Australian National University, Swinburne University of Technology, the University of Queensland, the University of Western Australia, the University of Melbourne, Curtin University of Technology, Monash University, and the Australian Astronomical Observatory. SkyMapper is owned and operated by The Australian National University’s Research School of Astronomy and Astrophysics. The survey data were processed and provided by the SkyMapper Team at ANU. The SkyMapper node of the All-Sky Virtual Observatory (ASVO) is hosted at the National Computational Infrastructure (NCI). Development and support of the SkyMapper node of the ASVO has been funded in part by Astronomy Australia Limited (AAL) and the Australian Government through the Commonwealth’s Education Investment Fund (EIF) and National Collaborative Research Infrastructure Strategy (NCRIS), particularly the National eResearch Collaboration Tools and Resources (NeCTAR) and the Australian National Data Service Projects (ANDS).
This research has made use of the NASA/IPAC Infrared Science Archive, which is funded by the National Aeronautics and Space Administration and operated by the California Institute of Technology.
Software packages used in this study include \texttt{Numpy} \citep{Numpy_2011}, \texttt{Matplotlib} \citep{Matplotlib_2007}.

\section*{Data Availability}
The SMSS data underlying this paper are available at the SkyMapper node of the All-Sky Virtual Observatory (ASVO), hosted at the National Computational Infrastructure (NCI) at \href{https://skymapper.anu.edu.au}{https://skymapper.anu.edu.au}. 6dFGS data are available at \href{http://www-wfau.roe.ac.uk/6dFGS/}{http://www-wfau.roe.ac.uk/6dFGS/} and the Final Data Release is available for public access. Data from NASA/ATLAS are publicly available at \href{https://fallingstar-data.com/forcedphot/}{https://fallingstar-data.com/forcedphot/}.




\bibliographystyle{mnras}
\bibliography{reference} 




\appendix
\section{Aperture effect on emission line ratios}\label{sec:cen_vs_wif}
\begin{figure}
    \centering
    \includegraphics{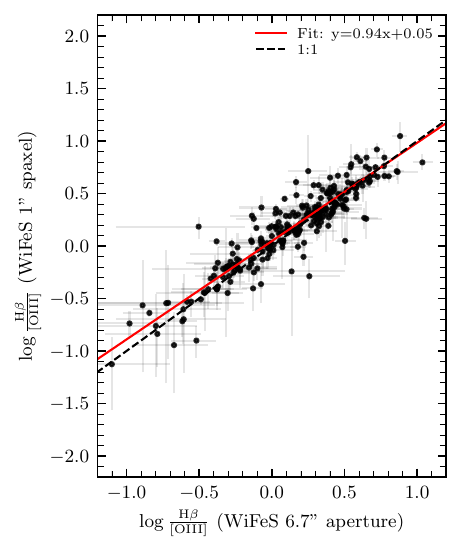}
    \caption{Comparison between \hb/[OIII] line ratios derived from WiFeS observations extracted from a single 1" spaxel targeting the central brightest pixel and from a 6dFGS-like 37 spaxel aperture, encompassing the extended host galaxy flux.}
    \label{fig:ratio_fit}
\end{figure}

The 6dFGS spectra are obtained from an aperture of 6.7", corresponding to an angular size of 12.4 kpc at $z=0.1$, thereby encompassing light from the extended host galaxy within the spectrum. The WiFeS integral field unit has a pixel scale of 1", resulting in an angular area approximately 37 times smaller than the 6dFGS aperture, thereby reducing the fraction of stellar light included the spectrum. Comparing different aperture extractions of the WiFeS spectra reveals that the single-spaxel spectra display a median increase of 12\% in the \hb/\oiii\ line ratios relative to the 6.7" aperture extractions, as shown in Figure \ref{fig:ratio_fit}.

\section{Light curves and spectra of CLAGN}\label{sec:all_cl_plots}
\begin{figure*}
    \centering
    \includegraphics[width=\linewidth]{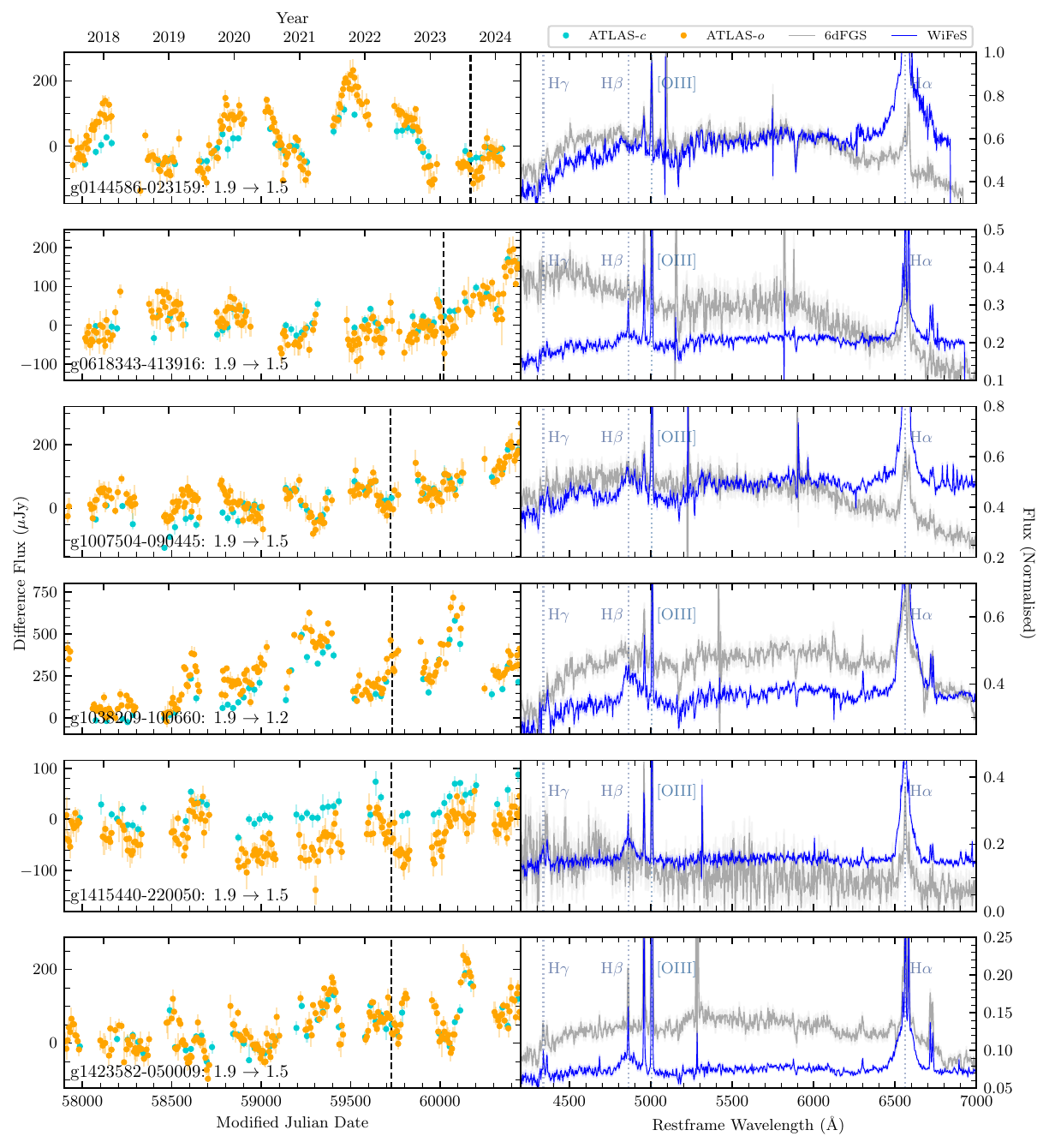}
    \caption{Turn on CLAGN identified in this work. See Figure \ref{fig:rep} for further explanations.}
    \label{fig:new_on_cls_1}
\end{figure*}
\begin{figure*}
    \centering
    \includegraphics[width=\linewidth]{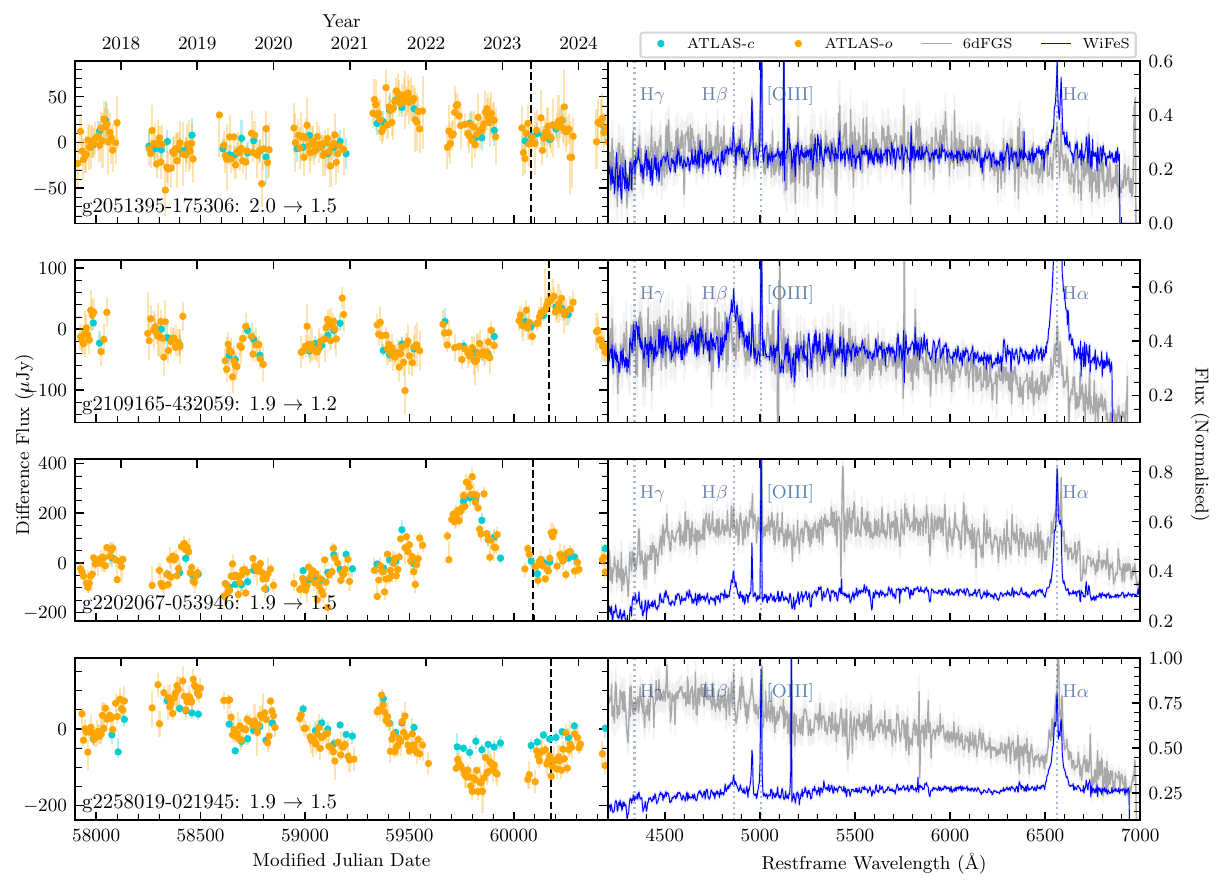}
    \caption{Turn on CLAGN identified in this work (continued from Figure \ref{fig:new_on_cls_1}).}
    \label{fig:new_on_cls_2}
\end{figure*}
\begin{figure*}
    \centering
    \includegraphics[width=\linewidth]{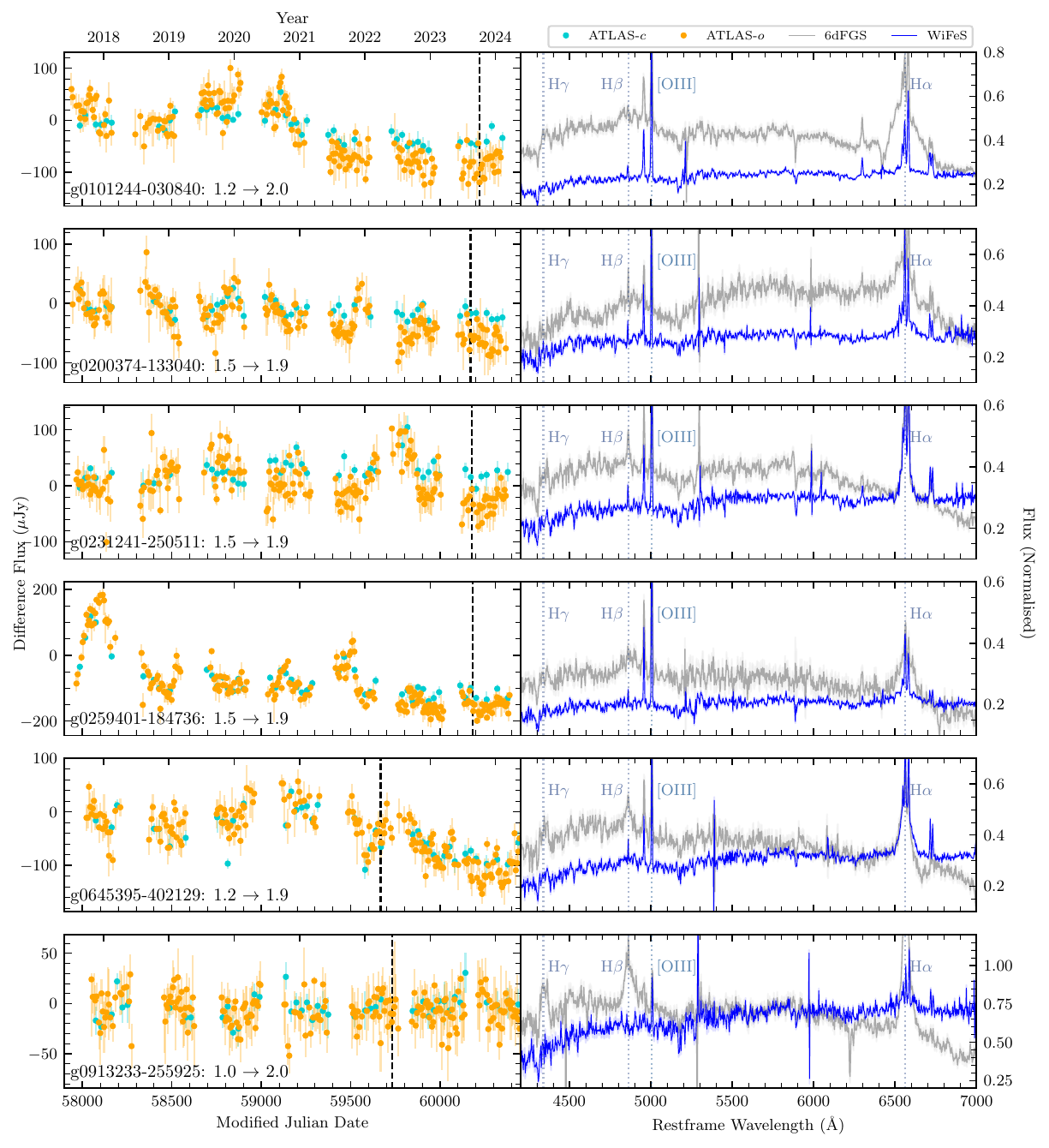}
    \caption{Turn off CLAGN identified in this work. See Figure \ref{fig:rep} for further explanations.}
    \label{fig:new_off_cls_1}
\end{figure*}
\begin{figure*}
    \centering
    \includegraphics[width=\linewidth]{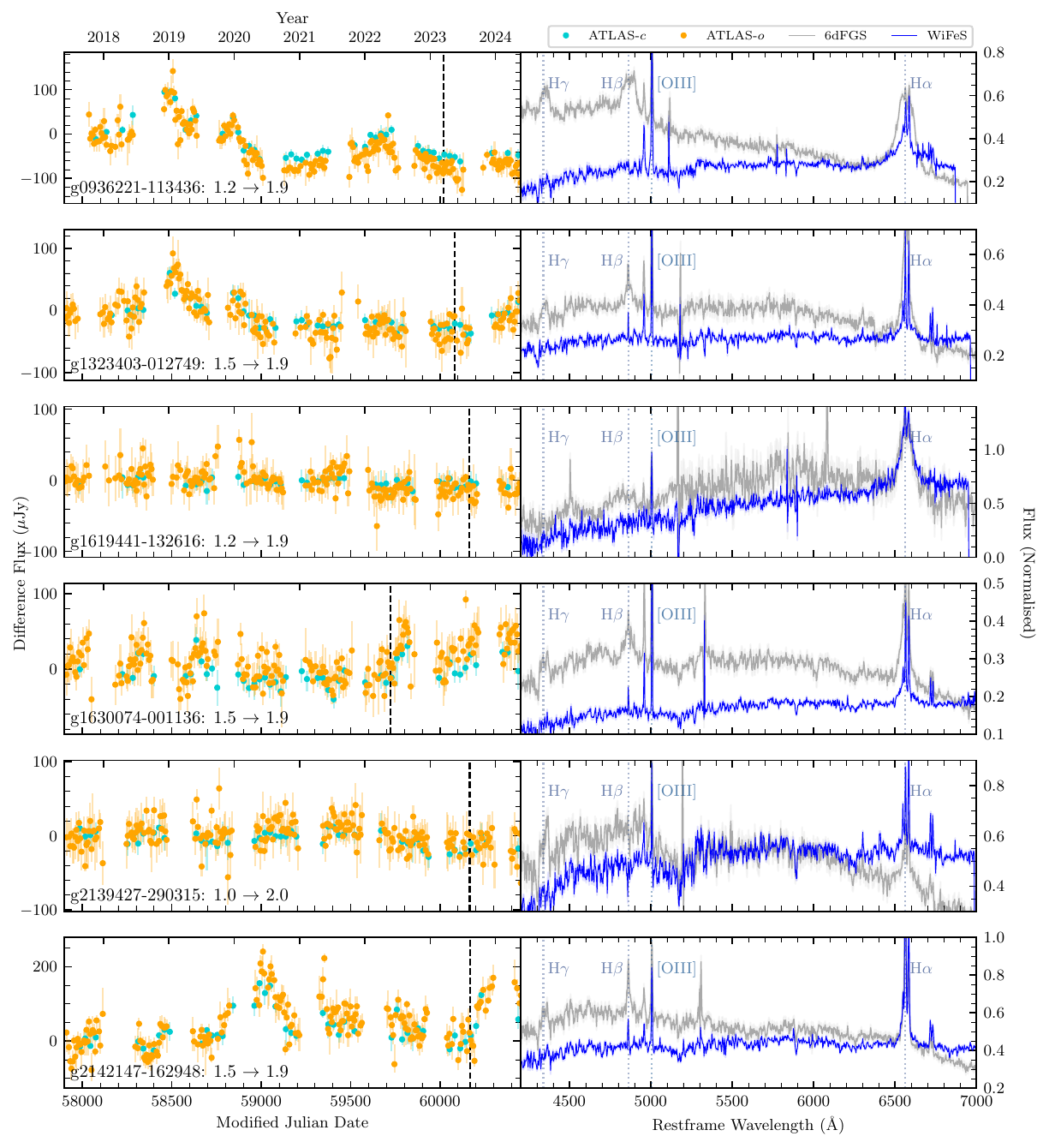}
    \caption{Turn off CLAGN identified in this work (continued from Figure \ref{fig:new_off_cls_1}). Note that 6dFGS object g1619441-132616 has $z=0$ contamination. This is not cross-talk as there is no $z=0$ object with adjacent 6dFGS spectral ID. The nearest {\it Gaia} source is 10.39" away. The contamination may be from a foreground Galactic HII region \citep[row 27;][]{sharpless_1959_hii_regions}.}
    \label{fig:new_off_cls_2}
\end{figure*}
\begin{figure*}
    \centering
    \includegraphics[width=\linewidth]{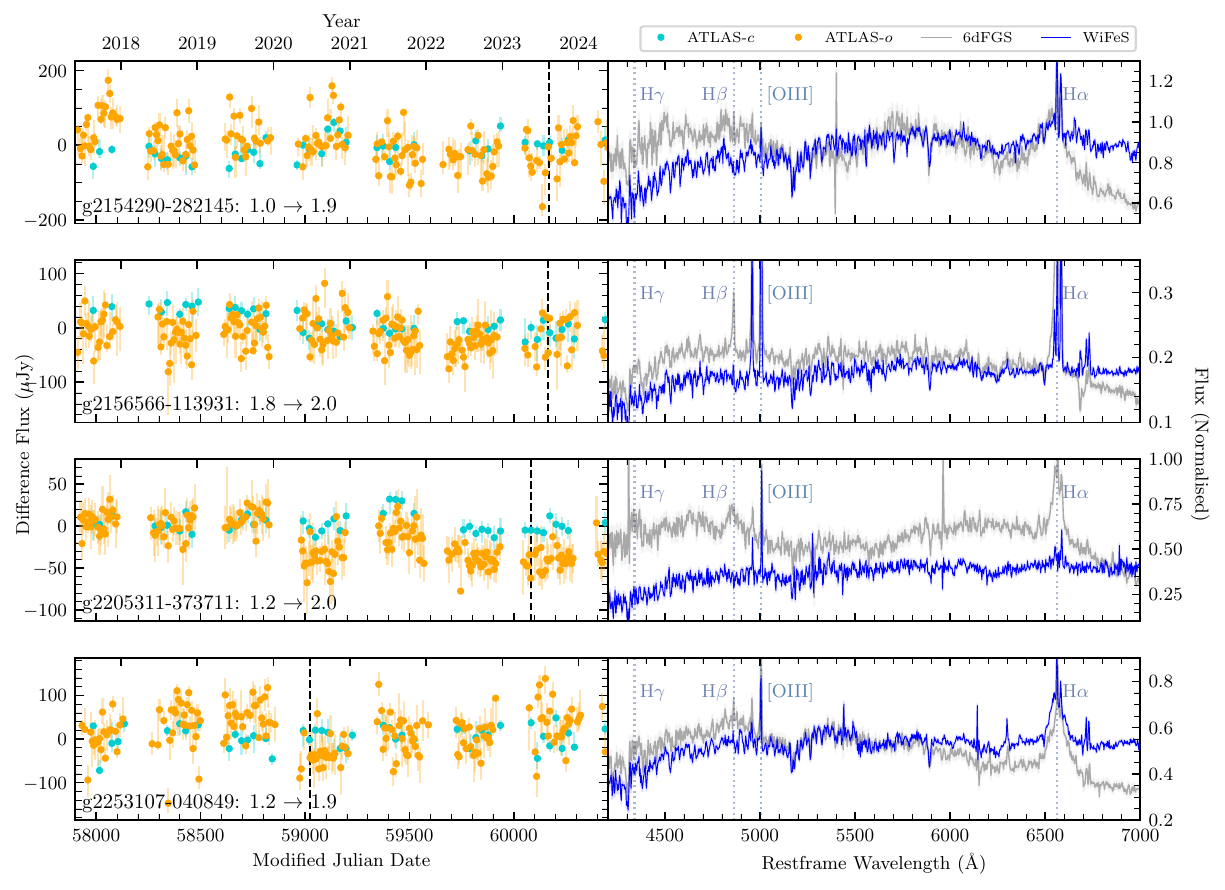}
    \caption{Turn off CLAGN identified in this work (continued from Figure \ref{fig:new_off_cls_2}).}
    \label{fig:new_off_cls_3}
\end{figure*}


\bsp	
\label{lastpage}
\end{document}